\documentclass[longauth]{aa}  

\usepackage{graphicx}
\usepackage{txfonts}
\usepackage{natbib}
\bibpunct{(}{)}{;}{a}{}{,} 

\newcommand{\hi}{H\,{\sc i}\xspace}

\newcommand{\HI}{H\,{\sc i}\xspace}
\newcommand{\kms}{km s$^{-1}$\xspace}
\newcommand{\msun}{$M_{\odot}$\xspace}

\begin{document}

\title{MeerKAT HI commissioning observations of MHONGOOSE galaxy ESO 302-G014}

\titlerunning{MHONGOOSE galaxy ESO302-G014}
\authorrunning{de Blok et al.}

   \author{W.J.G. de Blok
          \inst{1,2,3}
\and 
E. Athanassoula\inst{4}
\and 
A. Bosma\inst{4}
\and 
F. Combes\inst{5}
\and 
J. English\inst{6}
\and 
G.H. Heald\inst{7}
\and 
P. Kamphuis\inst{8}
\and 
B.S. Koribalski\inst{9,10,11}
\and 
G.R. Meurer\inst{12}
\and
J. Rom{\'a}n\inst{13}
\and 
A. Sardone\inst{14,15}
\and
L. Verdes-Montenegro\inst{13}
\and 
F. Bigiel\inst{16}
\and 
E. Brinks\inst{17}
\and
L. Chemin\inst{18}
\and 
F. Fraternali\inst{3}
\and 
T. Jarrett\inst{2}
\and 
D. Kleiner\inst{19}
\and 
F.M. Maccagni\inst{19}
\and 
D.J. Pisano\inst{20,21,22}
\and 
P. Serra\inst{19}
\and 
K. Spekkens\inst{23}
\and 
P. Amram\inst{4}
\and 
C. Carignan\inst{24,2}
\and 
R-J. Dettmar\inst{8}
\and 
B.K. Gibson\inst{25}
\and 
B.W. Holwerda\inst{26}
\and 
G.I.G J\'ozsa\inst{27,28,16}
\and 
D.M. Lucero\inst{29}
\and 
T.A. Oosterloo\inst{1,3}
\and 
A.J.T Ramaila\inst{27}
\and 
M. Ramatsoku\inst{28,19}
\and 
K. Sheth\inst{30}
\and 
F. Walter\inst{31}
\and 
O.I. Wong\inst{7,12,11}
\and 
A.A. Zijlstra\inst{32}
\and 
S. Bloemen\inst{33}
\and
P.J. Groot\inst{33,34,35}
\and
R. Le Poole\inst{36}
\and
M. Klein-Wolt\inst{33}
\and
E.G. K\"ording\inst{33}
\and
V.A. McBride\inst{37}
\and
K. Paterson\inst{38}
\and
D.L.A. Pieterse\inst{33}
\and
P. Vreeswijk\inst{33}
\and
P.A. Woudt\inst{34}
          }

   \institute{Netherlands Institute for Radio Astronomy (ASTRON), Oude Hoogeveensedijk 4, 7991 PD Dwingeloo, the Netherlands\\
              \email{blok@astron.nl}
         \and
             Dept.\ of Astronomy, Univ.\ of Cape Town, Private Bag X3, Rondebosch 7701, South Africa
         \and
             Kapteyn Astronomical Institute, University of Groningen, PO Box 800, 9700 AV Groningen, The Netherlands
    \and 
             Aix Marseille Univ, CNRS, CNES, LAM, Marseille, France
     \and 
             Observatoire de Paris, Coll\`ege de France, Universit\'e PSL, Sorbonne Universit\'e, CNRS, LERMA, Paris, France
     \and
             Department of Physics and Astronomy, University of Manitoba, Winnipeg, Manitoba, Canada, R3T 2N2, Canada
     \and 
             CSIRO Astronomy and Space Science, PO Box 1130, Bentley WA 6102, Australia
     \and
             Ruhr-Universit\"at Bochum, Faculty of Physics and Astronomy, Astronomical Institute, 44780 Bochum, Germany
     \and 
             Australia Telescope National Facility, CSIRO Astronomy and Space Science, P.O. Box 76, Epping, NSW 1710, Australia
     \and 
             Western Sydney University, Locked Bag 1797, Penrith South, NSW 1797, Australia
     \and
             ARC Centre of Excellence for All Sky Astrophysics in 3 Dimensions (ASTRO 3D)
     \and    
             International Centre for Radio Astronomy Research, The University of Western Australia, 35 Stirling Highway, Crawley, WA 6009, Australia
     \and
             Dep.to Astronomia Extragal\'actica Istituto Astrofisica de Andaluc{\'\i}a, Glorieta de la Astronomia s/n, 18008, Granada, Spain
     \and
             Department of Astronomy, The Ohio State University, 140 West 18th Avenue, Columbus, OH 43210, USA
     \and
             Center for Cosmology and Astroparticle Physics, 191 West Woodruff Avenue, Columbus, OH 43210, USA
     \and
             Argelander-Institut f\"ur Astronomie, Auf dem H\"ugel 71, 53121, Bonn, Germany
     \and
             Centre for Astrophysics Research, University of Hertfordshire, College Lane, Hatfield, AL10 9AB, UK
     \and 
             Centro de Astronom\'\i{}a, Universidad de Antofagasta, Avda. U. de Antofagasta 02800, Antofagasta, Chile
     \and 
              INAF - Osservatorio Astronomico di Cagliari, Via della Scienza 5, 09047, Selargius, CA, Italy
     \and   
              Department of Physics \& Astronomy, West Virginia University, Morgantown, WV 26506, USA
     \and
              Gravitational Waves and Cosmology Center, Chestnut Ridge Research Building, Morgantown, WV 26505, USA
    \and 
              Adjunct Astronomer, Green Bank Observatory, 155 Observatory Road, Green Bank, WV 24944, USA
    \and 
              Department of Physics and Space Science, Royal Military College of Canada, PO Box 17000, Station Forces, Kingston, Ontario K7K 7B4, Canada 
    \and     
              Laboratoire de Physique et de Chimie de l’Environnement, Observatoire d’Astrophysique de l’Universit\'e Ouaga I Pr Joseph KiZerbo (ODAUO), 03 BP 7021, Ouaga 03, Burkina Faso
    \and
              E.A. Milne Centre for Astrophysics, University of Hull, Hull, HU6 7RX, United Kingdom
    \and
                Department of Physics and Astronomy, 102 Natural Science Building, University of Louisville, Louisville, KY, 40292, USA
    \and 
             South African Radio Astronomy Oberservatory, Black River Park, 2 Fir Street, Observatory, Cape Town, 7925, South Africa
    \and
             Department of Physics and Electronics, Rhodes University, PO Box 94, Makhanda, 6140, South Africa
    \and 
             Department of Physics, Virginia Polytechnic Institute and State University, 50 West Campus Drive, Blacksburg, VA, 24061, USA
    \and 
             NASA Headquarters, 300 E Street SW, Washington, DC 20546, USA
    \and 
             Max Planck Institute for Astronomy, K\"onigstuhl 17, 69117 Heidelberg, Germany
    \and 
             Jodrell Bank Centre for Astrophysics, School of Physics and Astronomy, University of Manchester, Oxford Road, Manchester M13 9PL, UK
    \and 
             Department of Astrophysics/IMAPP, Radboud University, P.O. 9010,
            6500 GL, Nijmegen, The Netherlands
    \and     
            Inter-University Institute for Data Intensive Astronomy 
            \& Department of Astronomy, University of Cape Town, Private Bag X3,
            Rondebosch 7701, South Africa
    \and 
            South African Astronomical Observatory, P.O. Box 9, 7935
            Observatory, South Africa
    \and 
            Leiden Observatory, Leiden
            University, P.O. Box 9513, NL-2300 RA Leiden, The Netherlands
    \and
            IAU-Office For Astronomy for Development, P.O. Box 9, 7935
            Observatory, South Africa
    \and     
            Center for Interdisciplinary Exploration and Research in
            Astrophysics (CIERA), Northwestern University, 1800 Sherman Ave,
            Evanston, IL 60201, USA
      }

   \date{accepted by Astronomy \& Astrophysics, September 9, 2020}

 
  \abstract
    {}
    {We present the results of three
    commissioning \hi observations obtained with the MeerKAT radio
    telescope. These observations make up part of the preparation for the forthcoming
    MHONGOOSE nearby galaxy survey, which is a MeerKAT large survey project that
    will study the accretion of gas in galaxies and the link between
    gas and star formation.}
   {We used the available \hi data sets, along with ancillary data at other wavelengths, to study the morphology of the MHONGOOSE sample galaxy, ESO 302-G014, which is a nearby gas-rich dwarf galaxy.}
   {We find that ESO 302-G014 has a
   lopsided, asymmetric outer disc with a low column density.  In addition, we
   find a tail or filament of \hi clouds extending away from the
   galaxy, as well as an isolated \hi cloud some 20 kpc to the south
   of the galaxy. We suggest that these features indicate a minor
   interaction with a low-mass galaxy.  Optical imaging shows a
     possible dwarf galaxy near the tail, but based on the current data, we
     cannot confirm any association with ESO 302-G014.  Nonetheless, an
   interaction scenario with some kind of low-mass companion is still supported by the presence of a significant amount of
   molecular gas, which is almost equal to the stellar mass, and a number of
   prominent stellar clusters, which suggest recently triggered star
   formation.}
   {These
   data show that MeerKAT produces exquisite imaging data. The forthcoming full-depth survey
   observations of ESO 302-G014 and  other sample galaxies will,
   therefore, offer insights into the fate of neutral gas as it moves
   from the intergalactic medium onto galaxies.}


   \maketitle
%

\section{Introduction}

The relation between the gas content of a disc galaxy and the magnitude and
distribution of star formation is an important topic in the study of
galaxy evolution.  We can split this topic up into two components.

The first one pertains to the detailed physical conditions for in situ
disc gas to turn into stars. Much of this focuses on the conversion
from molecular gas to stars, but the atomic to molecular hydrogen
conversion is another key step in this process. The global relation
between gas density (atomic neutral hydrogen, or \hi, and molecular
gas) and star-formation rate (SFR), also known as the
Kennicutt-Schmidt law \citep{schmidt_1959,kennicutt_et_al_1998}, is
well-characterised. Comprehensive and detailed studies of the sub-kpc
relations have also become available in recent years. Some examples include studies based on The \HI Nearby Galaxy Survey
\citep[THINGS,][]{walter_et_al_2008} and its data \citep{bigiel_et_al_2008,
  leroy_et_al_2008}.  Other recent works include
\citet{zheng_et_al_2013} on the regulation of star formation;
as well as \citet{thilker_et_al_2007} and \citet{holwerda_et_al_2012} on the link
between star formation and \HI morphology (implying a physical link;
\citealt{heiner_et_al_2008}); studies of multi-scale star formation
and \hi \citep{elson_et_al_2019}; the link between SFR and gas based
on jellyfish galaxies \citep{jachym_et_al_2019,ramatsoku_et_al_2019,
  ramatsoku_et_al_2020, moretti_et_al_2020}; and
\citet{bacchini_et_al_2019} on the link between SFR and \hi volume
densities.

The second component deals with the origin of the neutral hydrogen
that turns into stars. In
many local disc galaxies the gas depletion time scale, which is the time
scale after which galaxies run out of molecular gas (at their
current SFRs), is only a few Gyr \citep{bigiel_et_al_2011}. The
molecular gas reservoir can be replenished from the neutral hydrogen
present in these galaxies, but we do not observe the
corresponding decline that is expected in the amount of cold neutral gas in galaxies
over cosmic time \citep[e.g.][]{noterdaeme_et_al_2012,
  kanekar_et_al_2016, hu_et_al_2019,sardone_et_al_2020}.

The conventional explanation is that galaxies acquire more gas over
their lifetime by accretion from the intergalactic medium
(IGM). However, in an inventory of possible \hi accretion features in
the   literature data available at the time, \citet{sancisi_et_al_2008} found that \hi
accretion can account for only $\sim 20$ percent of the gas supply needed to
sustain the current SFR in galaxies.  Most of the observations from
the literature used for their study had \hi column
density limits of $\sim 10^{20}$ cm$^{-2}$.  A natural 
explanation of this discrepancy is that accretion still occurs at lower
column densities, possibly in the form of partly ionised gas
filaments. The latter process has become known as ``cold accretion''
\citep{keres_et_al_2005,nelson_et_al_2013,nelson_et_al_2015,schaller_et_al_2015}.  Recent simulations highlight the importance
of accretion for galaxy evolution \citep{van_de_voort_et_al_2019}.

The Westerbork Hydrogen Accretion in LOcal GAlaxieS (HALOGAS) survey
\citep{heald_et_al_2011} was one of the first systematic attempts to
detect this \hi accretion and to put observational limits on the \hi
accretion rate at lower column densities ($\sim 10^{19}$ cm$^{-2}$) in
a sample of 24 gas-rich disc galaxies.  However, even with these more
sensitive data, the resulting \hi accretion rate is still
significantly lower than the observed SFR (Kamphuis et al., in
prep.). Most of the low-column density gas that is detected in HALOGAS
is extra-planar disc gas that most likely originates in the galactic
fountain process (\citealt{marasco_et_al_2019} and references therein;
also see \citealt{lucero_et_al_2015}).

Several possible explanations have been proposed for this lack of
visible accretion.  The first is that the external accretion process
can only, or mainly, be observed at low column densities well below
$10^{19}$ cm$^{-2}$. Simulations indicate that the distribution of \HI
should be much more extended at column densities around $\sim
10^{17-18}$ cm$^{-2}$ than at $10^{19-20}$ cm$^{-2}$
\citep[e.g.][]{popping_et_al_2009,van_de_voort_et_al_2019} and some of
this low column density \hi could make up the accreting material.  Ionised
gas at relatively low column densities is observed around star-forming
galaxies (e.g. \citealt{tumlinson_et_al_2017, werk_et_al_2013,
  werk_et_al_2014}) but its properties and origin are difficult to
establish from absorption measurements alone and, ideally, it  should
be seen in emission.

A second explanation -- which does not rule out the first  -- is
that direct cold accretion is not the dominant mechanism, but that accretion happens through
clouds that are launched from the disc as part of the galactic fountain
process.  These clouds
are launched from the disc into the hot halo, where they accrete gas which cools in their wake. This
gas is then brought back to the disc \citep{fraternali_et_al_2013,
  fraternali_2017}. 
  
Whatever the mechanism, the low-column density material is most
likely to be clumpy on smaller ($\sim$ kpc) scales
\citep{popping_et_al_2009, wolfe_et_al_2016}, so for the next
observational step, both good column density sensitivity and spatial
resolution are needed. The MHONGOOSE survey (MeerKAT \hi Observations
of Nearby Galactic Objects: Observing Southern Emitters;
\citealt{de_blok_et_al_2016}) with the MeerKAT telescope will have
both of these. It will probe the resolved distribution of \hi at
uniquely deep column densities, allowing us to reveal what is possibly
accreting cold gas and quantify its contribution compared to the recycled
gas from the fountain process, thereby further opening up the observational
parameter space to probe the accretion of gas in galaxies.

The MeerKAT radio synthesis telescope \citep{jonas_et_al_2016,camilo_2018,mauch_et_al_2020} is a
64-dish interferometer located in the Karoo desert in South Africa. Each dish
has a Gregorian design with an effective 13.5m diameter. MeerKAT is a
precursor telescope of the future SKA1-MID telescope \citep{braun_et_al_2015}. It will
eventually also form part of this telescope.

Much of the MeerKAT science case is addressed in a number of Large
Survey Projects (LSPs) which are multi-year observing projects with
$>1000$ hours of observing
time apiece. MHONGOOSE \citep{de_blok_et_al_2016} is an LSP specifically
designed to provide both high-resolution and high-sensitivity observations of
nearby galaxies, which is handy in addressing the two main components of the link
between gas content and star formation described above.

A full description of the MHONGOOSE survey, the sample, and the
  sample selection will be given in a future paper.  Here, we give a
  brief summary of the essentials only. MHONGOOSE will
observe 30 disc and dwarf galaxies over a total of 1650 hours (55
hours per galaxy) over the next five years. The sample covers five
bins in logarithmic \hi mass in the range $6<\log(M_{\rm HI}/M_\odot)
< 11$ with an equal number of galaxies in each bin. These galaxies
were selected from the Survey for Ionization in Neutral Gas Galaxies
(SINGG) \citep{meurer_et_al_2006} sample. SINGG's main goal is to
provide a census of star formation in \hi-selected galaxies using
H$\alpha$ and $R$-band observations.  Galaxies with specific
inclinations and SFRs were selected to optimise the modelling and
analysis and to ensure a comprehensive coverage of the \hi mass and
SFR distribution. For completeness, we list the sample and some
  basic properties of these galaxies in Table \ref{tab:sample} in the
  Appendix, but, as noted, a full description will be given in a future
  paper.

The expected \hi column density sensitivity of the MHONGOOSE observations is
$\sim 3 \cdot 10^{18}$ cm$^{-2}$ at $3\sigma$ over a 16 \kms channel
at $30''$ resolution. This implies that at $\sim 1'$ resolution the
column density sensitivity will be equal to those of the deepest
Parkes or Green Bank telescope (GBT) \hi observations of the MHONGOOSE
galaxies \citep{sorgho_et_al_2019, sardone_et_al_2020}, but at an
angular resolution an order of magnitude finer (cf.\ Fig.\ 5 in
\citealt{de_blok_et_al_2016}).

Similarly at $\sim 7''$, the highest resolution we can achieve
for \HI in MHONGOOSE, the column density sensitivity will be a factor of four better ($\sim 6
\cdot 10^{19}$ cm$^{-2}$ at $3\sigma$ over a 16 \kms channel at $7''$) than
the THINGS observations \citep{walter_et_al_2008}, enabling the
mapping of \HI beyond the star-forming disc at these very high
resolutions.

The spectral resolution of the final observations will be 3.3 kHz, or
0.7 \kms at the \hi rest frequency of 1420.4 MHz, using the narrow-band mode of the MeerKAT correlator, dividing the
bandwidth of 107 MHz available in that mode into 32768 channels. This
will enable highly accurate observations of the \hi profiles, with the
ability to separate them into cold and warm components, thus tracing the
different neutral hydrogen components of the interstellar medium
(ISM).

As part of the MeerKAT telescope commissioning, one of the MHONGOOSE
galaxies, ESO 302-G014, was observed multiple times over the course of
two years at various stages of completeness of the array and the
correlator. The initial choice for this galaxy was its observability
at night during the first observations and the  lack of
high-quality resolved \hi data in telescope archives at the time.

In this paper, we describe three MeerKAT commissioning \hi data sets from
ESO 302-G014. We also present a brief discussion of the morphology of
the galaxy and possible scenarios for its evolution.  Section 2
describes the properties of the galaxy. Section 3 gives details on the
observations and data reduction. Section 4 describes the data and
presents a first analysis. Section 5 presents a short discussion. 
We summarise our work in Section 6, including our expectations for the full
survey. Unless noted otherwise, all velocities mentioned in this paper
are heliocentric.

\section{ESO 302-G014\label{sec:auxdata}}

\begin{figure}
  \centering\includegraphics[width=0.95\columnwidth]{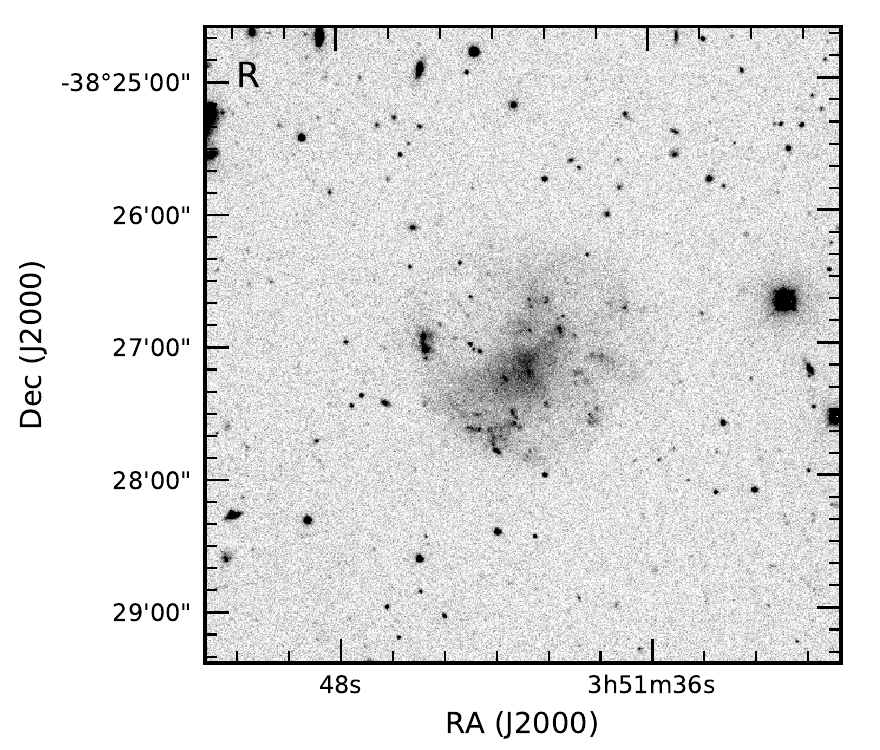}
  \caption{SINGG $R$-band image of ESO 302-G014
    \citep{meurer_et_al_2006}. The star clusters discussed in the text
    are visible as the two brightest sources $45''$ east of the center
    of the galaxy. The limiting surface brightness of this image is 27.6 AB mag arcsec$^{-2}$.}
    \label{fig:singg}
\end{figure}

ESO 302-G014 (HIPASS J0351--38) is a low-mass, late-type and gas-rich
dwarf galaxy.  The SINGG $R$-band image of ESO 302-G014 is shown in
Fig.~\ref{fig:singg}. The galaxy has an absolute magnitude of $M_B =
-15.33$ \citep{prieto_et_al_2012}, making it somewhat less luminous
than the SMC.  As projected on the sky, it is located between the Fornax cluster and the Dorado
group \citep{venhola_et_al_2018}, but it is  not itself part of any group
\citep{kourkchi_et_al_2017}.  The galaxy has a recessional velocity of
$871$ \kms, translating into an assumed distance of 11.7 Mpc, as
derived in \citet{meurer_et_al_2006} from the heliocentric velocity
using the model by \citet{mould_et_al_2000}. At this distance, $1''$
corresponds to 57 pc and $1'$ to 3.4 kpc.

It is the host galaxy of SN2008jb, a core-collapse Type II supernova
extensively studied by \citet{prieto_et_al_2012}. They find that the
supernova occurred in an actively star-forming part of the galaxy,
dominated by two young stellar clusters (visible as the two brightest
sources $\sim 45''$ to the east of the center of the galaxy in
Fig.~\ref{fig:singg}). Various estimates lead them to conclude that
the time scale of the star formation in this area is $\sim 9$ Myr, which
would imply a high-mass progenitor for the supernova.

The star-formation rate based on SINGG H$\alpha$ data \citep{meurer_et_al_2006} (for an assumed initial
mass function with a \citealt{salpeter_1955} slope and spanning the
mass range 0.1 to 100 M$_\odot$) is $\sim 0.03$ \msun yr$^{-1}$
(G.\ Meurer, priv. comm.). This agrees reasonably well with the SFR derived from
ultra-violet GALEX data \citep{meurer_et_al_2009,lee_et_al_2011}.

These numbers, combined with a stellar mass estimate of $4 \cdot 10^7$
\msun \citep{prieto_et_al_2012} and an \hi mass of $3 \cdot 10^8$ \msun \citep[the \hi
  mass is discussed extensively in the current
  paper]{meurer_et_al_2006} lead to SFR values per unit stellar and
\hi mass of $\log({\rm SFR}/M_\star) \simeq -9.1$ [yr$^{-1}$] and $\log({\rm
  SFR}/M_{\rm HI}) \simeq -9.9$ [yr$^{-1}$], which are typical for local
star-forming dwarf galaxies (cf.\ \citealt{lee_et_al_2011}).

An independent stellar mass determination using WISE data confirms
this picture.  Based on the W1 flux and the relations from
\citet{cluver_et_al_2014}, we find a value of $\log(M_\star/M_{\odot})
= 7.82 \pm 1.82$ or $M_\star \sim 6.6 \cdot 10^7M_{\odot}$. The
Spitzer S4G data report a stellar mass of ~$1.1 \cdot 10^8M_{\odot}$,
corrected to the distance adopted here;
compare with \citet{munoz_mateos_et_al_2015}.

These numbers, again, lead to a specific SFR that is normal for dwarf
galaxies. They do show, however, that the galaxy is very gas
rich. Depending on the stellar mass value adopted, the mass ratio
$M_{\rm HI}/M_{\star}$ varies between $\sim 2.9$ and $\sim 7.5$.

The occurrence of SN2008jb and the relative (projected) proximity to
the Fornax cluster has led to the availability of much archival data
for this galaxy, including observations by ALMA and HST. As these data
sets are used later in this paper, we briefly list their technical
details here.

The ALMA 12-m total power (TP) array observations were carried out as
part of a programme to observe 64 Fornax galaxies in band 3, during
cycle 5 as part of the project 2017.1.00129.S (PI K.\ Morokuma). For ESO
302-G014, J0406-3826 was observed as a calibrator.  The CO($J=1-0$)
emission line was included in the highest frequency USB spectral
window (SPW17), while the three others served as continuum.  The TP
on-the-fly observations were carried out for 2$^h$45$^m$ in May 2018
with three 12m-antennae. The angular resolution of the TP observations
is $54''$, and the velocity resolution 2.9 \kms.  The data were
calibrated with the Common Astronomy Software Applications (CASA,
\citealt{mcmullin_et_al_2007}). A data cube was produced spanning
$212'' \times 212''$, and 4800 \kms. The noise level is 0.16 Jy beam$^{-1}$ in channels of 10
\kms.  The emission is marginally resolved and an integrated CO
spectrum was produced by integrating over an area of $130'' \times
130''$ centered on the optical position.

The HST data consist of a ACS/WFC3 pointing in four filters (F606W, F336W,
F657N, F160W) obtained as follow-up of SN2008jb 
as part of Proposal 12992 (P.I.\ J.\ Prieto) in October 2012. Reduced
data were retrieved from the Hubble Legacy Archive.

In addition, we used data from the Dark Energy Camera Legacy
  Survey (DECaLS; \citealt{dey_et_al_2019}). The $g$, $r,$ and $z$
  images have surface brightness limits of respectively 29.2, 29.0,
  and 27.5 mag arcsec$^{-2}$ (3$\sigma$ in $10'' \times 10''$ boxes) in
  the area of ESO 302-G014.
These ancillary data sets are all discussed in Section~\ref{sec:interaction}.

\section{MeerKAT observations and data reduction}

The first commissioning observations of ESO 302-G014 were done in
early 2018 with the MeerKAT-16 array.  This early commissioning array
consisted of 16 MeerKAT dishes with short baselines. These limited-resolution data
showed interesting features in this galaxy (described below),
compelling us to target it in subsequent commissioning
stages with the full MeerKAT-64 array as described in this
paper. These further observations consisted of an observation in 2019
using the final MeerKAT SKARAB correlator in its 4k mode (channel
width of 44 \kms), and an observation in early 2020 with the full
array using the 32k mode (channel width of 5.5 \kms). The final
MHONGOOSE observations will have a velocity resolution that is another
factor of eight better, giving a
channel width of 0.7 \kms. The data described below were not Hanning-smoothed.

\subsection{MeerKAT-16 observation}

ESO~302-G014 was observed in January 2018 with the MeerKAT-16 array.
The standard calibrator B1934-638 was used as the flux calibrator,
J0408-6548 was used as the bandpass calibrator, and B0451-282 as the gain
calibrator.  The flux and bandpass calibrators were observed for five
minutes each every two hours, with the gain calibrator observed for two
minutes every ten minutes.  The (now decommissioned) ROACH2 correlator
was used, with a channel width of 26.1 kHz (5.5 \kms at 1420.4 MHz).
Further details of the observation are given in
Table~\ref{tab:observations}.
 
The data were reduced using a pre-release version of the Containerized
Automated Radio Astronomy Calibration (CARACal) pipeline
(\citealt{jozsa_et_al_2020}; see also
\citealt{serra_et_al_2019}). This is a containerised pipeline which
allows for the automatic running of the relevant components from different
radio astronomy packages as part of one single scripted execution.
The usual steps of flagging data, cross-calibration, splitting of
data, self-calibration, continuum subtraction, as well as line imaging
and deconvolution can all be done with a single call to the
pipeline. Furthermore, the pipeline provides diagnostic plots that can
be used to assess the quality of the data and the calibration and
reduction.  Especially in the early stages of the reduction, the
  pipeline can be run in stages and on subsets of the data so
  that diagnostic plots and output products can be inspected and if
  necessary, pipeline parameters can be adjusted before a full run on
  the complete data set.

Using the pipeline, continuum was subtracted in two stages. Firstly, the
final self-calibration continuum model was subtracted and, afterwards, any
remaining residual continuum was subtracted in the $uv$-plane using a
first-order polynomial fit to adjacent line-free channels.

The pipeline then produced the final data cube using a
two-step deconvolution process where the output of a first, shallow
deconvolution run was used as a mask for the second, deeper
deconvolution run.  Further details on this data set are given in
Table \ref{tab:noises}. We refer to this observation as the
`MeerKAT-16' data.

\subsection{MeerKAT-64 4k observation}

ESO 302-G014 was observed again in July 2019 with the full MeerKAT array,
with 61 dishes participating in the observation. The L-band receivers
were used, covering a frequency range from approximately 900--1670
MHz. The SKARAB correlator in 4k mode was used, dividing the band
into 4096 channels with a channel width of 208.9 kHz each, or $\sim 44$
\kms. Observations were taken in full polarization mode.

The standard calibrator J0408-6544 was observed as the flux and bandpass
calibrator, with J0440-4333 as the gain calibrator. The target and
calibrators were observed sequentially, in a cycle of 10 min on
target, 5 min on J0408-6544, and 2 min on J0440-4333. The total time
spent on the flux and bandpass calibrator is longer than usual, but this
allowed for an accurate study of the various calibration steps for
commissioning purposes. Further details of the observation are listed
in Table~\ref{tab:observations}.

\begin{table*}
    \caption[]{Observation details}
    \label{tab:observations}
    \centering
    \begin{tabular}{l l l l}
      \hline
      \hline
        {} & MeerKAT-16 32k & MeerKAT-64 4k & MeerKAT-64 32k\\
      \hline
        Observing date (UTC) & 28 Jan 2018 & 21 July 2019 & 07 Jan 2020\\
        Observing time (UTC) & 12$^h$03$^m$--23$^h$23$^m$ & 01$^h$06$^m$--12$^h$47$^m$ &  15$^h$40$^m$--21$^h$53$^m$\\
        Array configuration & 16 dishes\tablefootmark{a} & 61 dishes\tablefootmark{b} & 59 dishes\tablefootmark{c}\\
        Total observing time & 680 min & 620 min & 373 min\\
        Total time on target & 504 min & 340 min & 258 min\\
        Channel width & 26.1 kHz & 208.9 kHz & 26.1 kHz\\
        Number of channels & 32768 & 4096 & 32768\\
        \hline
    \end{tabular}
    \tablefoot{ \tablefoottext{a}{Antennas m000, m006, m011, m012,
        m015, m016, m021, m023, m024, m027, m029, m031, m033, m034,
        m035, m039.}  \tablefoottext{b}{Does not include antennas m001, m025,
        m039.}  \tablefoottext{c}{Does not include antennas m026, m028, m032,
        m051, m063.}
    }
\end{table*}

The data were again reduced using a pre-release version of the CARACal
pipeline. After delay, bandpass and gain calibration solutions were
derived, three rounds of self-calibration were then applied to the
data. A continuum image was created using the central 100 channels
($20.9$ MHz centered on 1416.5 MHz), using a Briggs robustness
weighting parameter \citep{briggs_1995} of $r=0.0$. The noise in this
image is 14 $\mu$Jy with a beam size of $8.2'' \times 7.1''$. The most
prominent feature in the continuum image at this angular resolution is
a peak at the location of SN2008jb. We find a peak brightness at that
location of $96.8\,\mu\mathrm{Jy\,beam^{-1}}$. The continuum image is
discussed further in Sect.\ \ref{sec:SF}.

The continuum emission was removed using the source list derived from
the clean model of the final self-calibrated continuum image followed
by a linear fit using the channels straddling the \hi emission, and a
subsequent subtraction in the $uv$ plane.  We extracted the central
eight channels that were then deconvolved in a two-step process as
described above.

\begin{table*}
    \caption[]{Noise levels for the \hi cubes}
    \label{tab:noises}
    \centering
    \begin{tabular}{l r l r r r r r l l }
      \hline
      \hline
      \noalign{\vskip 2pt}
      array & corr. &  $r$  & noise $\sigma$ & beam  & PA (beam) &$N_{\rm HI}$ & $N_{\rm HI}^{3\sigma}$&  $F$ & $M_{\rm HI}$ \\
      \noalign{\vskip 2pt}
      & mode  &       & $(\mu{\rm Jy\, beam}^{-1})$ & (arcsec) & $(^\circ)$ & \multicolumn{2}{c}{$(10^{18}$ cm$^{-2}$)} & (Jy\,\kms) & $(10^8 \,M_{\odot})$\\
      (1) & (2) & (3) & (4) & (5) & (6) & (7) & (8) & (9) & (10) \\
      \hline
      \noalign{\vskip 2pt}
      MeerKAT-16 & 32k  & 2.0 & 1040 & $62.7 \times 54.2$ & 68.4& 1.9 & 9.7 & $10.93 \pm 0.29$ & $3.5 \pm 0.1$\\
      \hline
      \noalign{\vskip 2pt}
      MeerKAT-64 & 4k  & 0.0 & 120 & $8.0 \times \phantom{0}7.5$   & 97.5& 97.8 & 58.7\rlap{\tablefootmark{a}}  & 10.70  &  3.5\\
      &4k  & 0.5 & 97  & $11.5 \times 10.1$ & 113.2&40.8 & 24.5\rlap{\tablefootmark{a}}  & 11.69  &  3.8\\
      &4k  & 1.0 & 86  & $23.7 \times 16.6$ & 124.6&10.7 & 6.4\rlap{\tablefootmark{a}}  & 12.14  &  3.9\\
      &4k  & 1.5 & 85  & $30.7 \times 21.8$ & 122.9&6.2  & 3.7\rlap{\tablefootmark{a}}  &11.93  &  3.9\\
      \hline
      \noalign{\vskip 2pt}
      MeerKAT-64 &32k & 0.0 & 358 & $8.2 \times \phantom{0}6.8$ &155.5& 39.3 & 201.2 & $10.45 \pm 0.16$   &  $3.4 \pm 0.1$\\
      &32k & 0.5 & 283 & $13.7 \times \phantom{0}9.8$  & 139.7 &12.9 & 66.0 & $11.28 \pm 0.14$  &  $3.6 \pm 0.1$ \\
      &32k & 1.0 & 255 & $25.4 \times 18.7$ &139.4& 3.3  & 16.9 & $11.98 \pm 0.11$  &  $3.9 \pm 0.1$ \\
      &32k & 1.5 & 247 & $30.3 \times 25.9$ &143.1& 1.9  & 9.7 & $11.87 \pm 0.09$  &  $3.8 \pm 0.1$\\
      \hline
    \end{tabular}
    \tablefoot{(1): MeerKAT array configuration. (2) Correlator
      mode. The 4k mode has a channel width of 44 \kms; the 32k mode 5.5 \kms. (3) Value of the robustness parameter $r$. (4) Noise $\sigma$
      in a single channel in $\mu$Jy beam$^{-1}$. (5) Size of the synthesised beam
      in arcseconds. (6) Position angle of the major axis of the beam. (7) The $1\sigma$ column density sensitivity in a
      single channel ($\times 10^{18}$ cm$^{-2}$). (8)  The 3$\sigma$ column density for a 16 \kms line width (assuming a single channel). (9) Integrated \HI
      flux $F$ (in Jy \kms) with associated uncertainty. No uncertainty is given for the 4k data, due to the small number of channels with emission. (10) Total \HI mass, derived from column (9)
      and assuming a distance of 11.7 Mpc ($\times 10^8\ M_\odot$).
      \tablefoottext{a}{These column densities are scaled from values measured with a wider 45 \kms channel and are  listed here merely for comparison. They cannot be measured directly from the data set.}}
\end{table*}

We produced a number of data cubes this way, using various values of
the robust parameter $r$ to explore the image quality
and the \HI morphology at various resolutions. Table \ref{tab:noises}
lists the resulting synthesised beam values, as well as noise levels
and column density limits. The latter is calculated as a 1$\sigma$
limit for a single 209 kHz channel. In addition, we also show the
$3\sigma$ column density limits for a fixed line width to facilitate
direct comparison with the other two data sets.

The natural-weighted data set has a noise per channel of 85
$\mu$Jy beam$^{-1}$. We calculate an expected
thermal noise of 79 $\mu$Jy beam$^{-1}$, showing that the noise levels in the data
are close to the expected value. In the rest of this paper, we
 refer to this dataset as the `4k' data.

\subsection{MeerKAT-64 32k observation}

Our final ESO 302-G014 commissioning observation took place in January
2020, this time as part of commissioning of the 32k mode of the SKARAB
correlator. This covers a total bandwidth of 856 MHz with 32768
channels that are each 26.1 kHz wide. This corresponds to 5.5 \kms at 1.4
GHz. The full array was used, with 59 active dishes.

The standard calibrator J0408-6544 was observed as the primary flux and
bandpass calibrator, J0440-4333 as the gain
calibrator. The target and phase calibrator were observed sequentially, in
a cycle of 10 min on target, and 2 min on the secondary
calibrator. The primary calibrator was observed twice during the
observation for a total of 20 mins. Additional observing parameters
are listed in Table~\ref{tab:observations}.

The reduction of the data was identical to that described above,
except that (in order to speed up the reduction process) we ran the
self-calibration step on a smaller measurement set created by binning
the data in bins of 40 channels wide.  The resulting self-calibration
solutions were then interpolated and applied to the measurement set
with the original spectral resolution. Continuum subtraction and
creation and deconvolution of data cubes then proceeded as above.

The cube produced with a robust value $r=1.0$ (close to
natural-weighted) has a noise per channel of 255 $\mu$Jy
beam$^{-1}$. The expected thermal noise is 226 $\mu$Jy beam$^{-1}$,
again showing good agreement. In the rest of this paper, we  refer
to this dataset as the `32k' data.

\section{Description of the HI data\label{sec:description}}

In Table \ref{tab:noises}, we list the noise levels per channel for the
MeerKAT-16 as well as the MeerKAT-64 4k and 32k data, using different
robust values, $r$. Also listed are the resulting beam sizes and
1$\sigma$ single-channel column-density sensitivities.

\begin{figure}
  \centering\includegraphics[width=0.8\columnwidth]{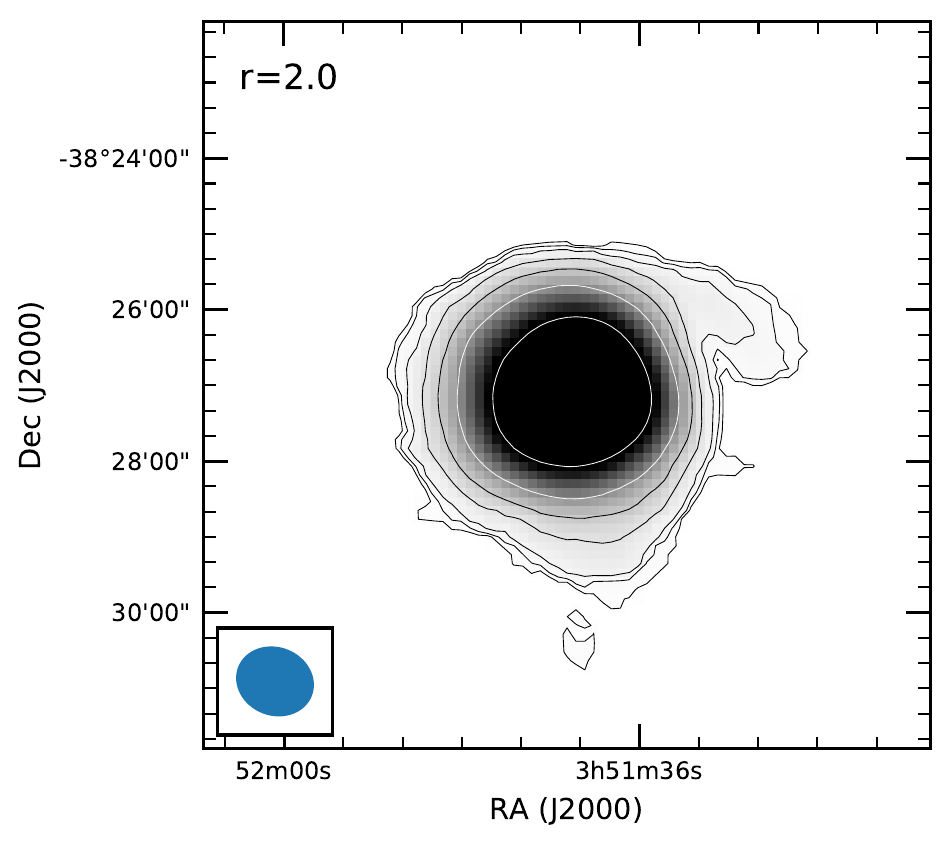}
  \caption{Integrated \hi map (zeroth moment) of the MeerKAT-16 data
    using $r=2.0$.  Contour levels are (0.05, 0.1, 0.2, 0.5, 1.0, 2.0,
    5.0, 10.0) $\cdot 10^{20}$ cm$^{-2}$. The MeerKAT-16 beam is shown
    in the bottom-left corner. We note the extension visible in the
    north-western part of the disc.
    \label{fig:mk16}}
\end{figure}

For the MeerKAT-16 data, we do not show the channel maps due to the
limited angular resolution of the data, which, of course, is massively
improved by the MeerKAT-64 data. Instead, we show in
Fig.~\ref{fig:mk16}, the integrated \hi or zeroth-moment map, as this
was the motivation to use this galaxy for further commissioning observations.

\begin{figure*}
  \centering\includegraphics[width=0.95\textwidth]{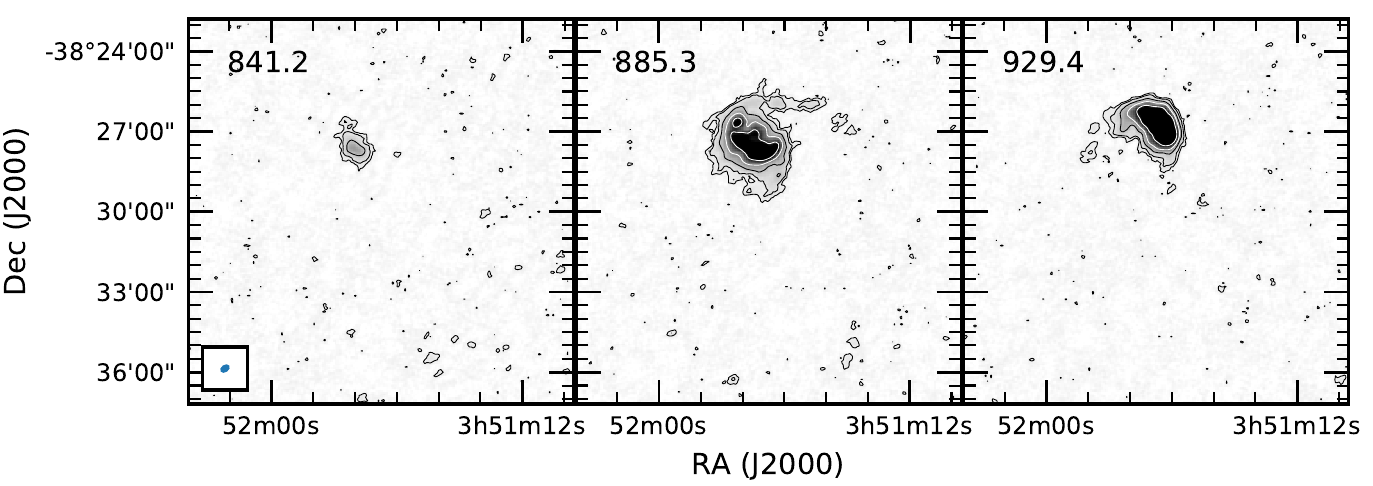}
  \caption{Channel maps of the 4k data containing \HI emission produced using
    $r=1.0$. The velocity of each channel is given in the top left in
    \kms. The beam (as given in Table \ref{tab:noises}) is indicated in the box
    at the lower left of the left-hand panel. The contour levels are drawn at (2.5, 5.0,
    12.5)$\sigma$ for the black contours, and at (25, 50)$\sigma$ for
    the white contours. The noise $\sigma$ equals 86 $\mu$Jy beam$^{-1}$.}
    \label{fig:chanmaps4k}
\end{figure*}

The map was created using the SoFiA source finder
\citep{serra_et_al_2015}. We  searched for emission brighter than
$3\sigma$ using all four combinations of the original spatial and
velocity resolution, a Gaussian spatial smoothing kernel equal to the
synthesised beam, and a three-channel boxcar velocity smoothing kernel. The
built-in comparison between positive and negative detections with a
reliability threshold of 0.90 was then used to distinguish between
noise peaks and real emission.

In Fig.~\ref{fig:mk16}, despite the limited resolution, we clearly see
a filament-like structure emanating from the northern part of
the disc and extending to the west-south-west. This intriguing
feature led us to use ESO 302-G014 for future commissioning
observations as well. We discuss the feature in more detail below.

In Fig.~\ref{fig:chanmaps4k}, we present the three channel maps
containing \HI emission from the 4k data created using a robustness
value of $r=0.5$.  Most of the emission is concentrated in the central
channel, which is consistent with the velocity width of the galaxy and the
large channel width. The central velocity channel clearly shows the
extension to the west, along with some possible faint clouds at $\sim
7.5'$ south (around $(\alpha, \delta) = 3^h51^m24^s, -38^\circ
35'30''$).

\begin{figure*}
  \centering\includegraphics[width=0.90\textwidth]{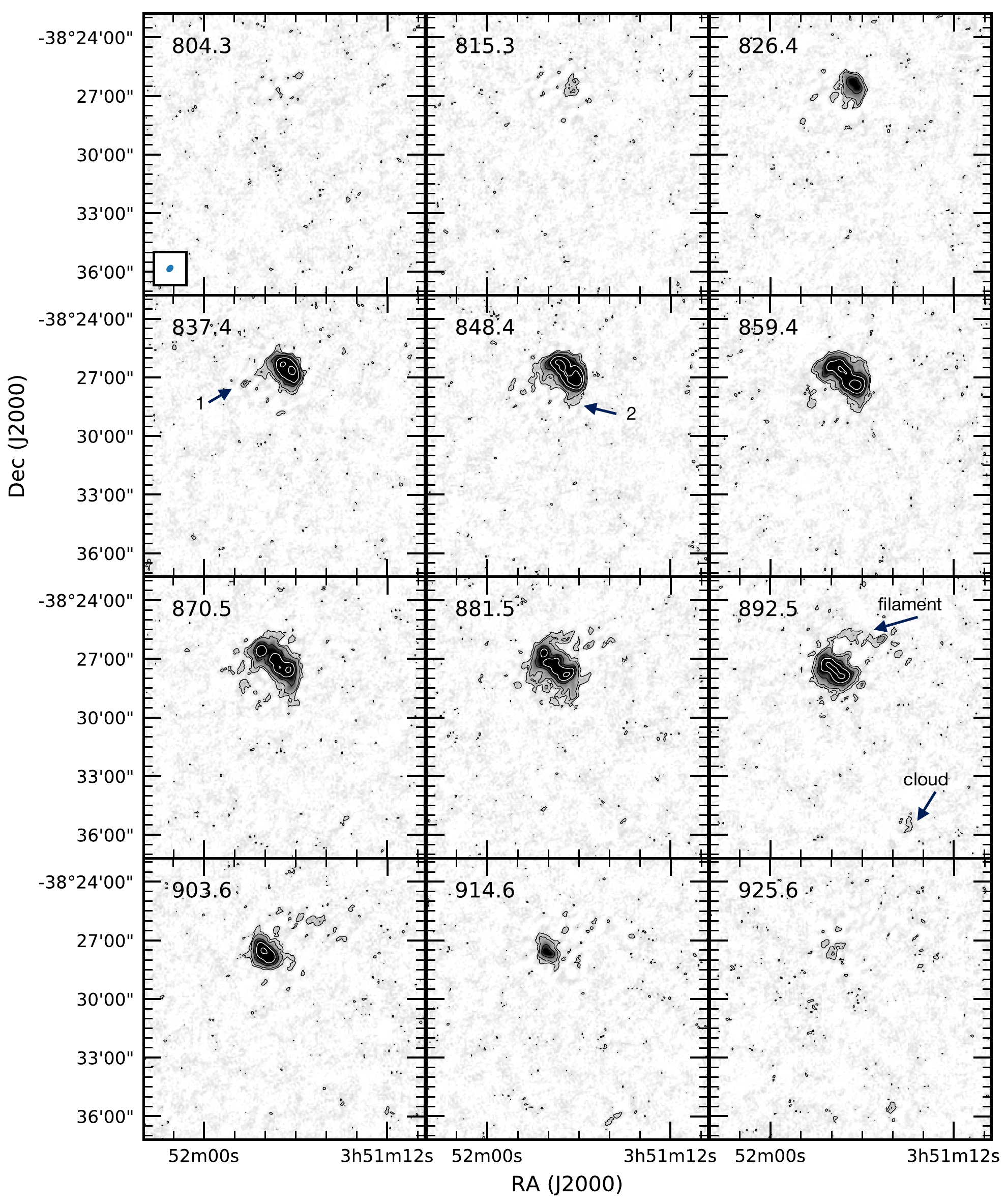}
  \caption{Channel maps of the 32k data containing \HI emission
    produced using $r=1.0$. Only every second channel is shown. The
    velocity of each channel is given in the top left in \kms. The
    beam (as given in Table \ref{tab:noises}) is indicated in the box
    in the top-left panel. The contour levels are drawn at (2.5, 5.0,
    12.5)$\sigma$ for the black contours, and at (25, 50)$\sigma$ for
    the white contours. The noise $\sigma$ equals 255 $\mu$Jy
    beam$^{-1}$. The cloud and filament are indicated. Numbers ``1''
    and ``2'' refer to features described in the text.}
    \label{fig:chanmaps32k}
\end{figure*}

The same velocity range is shown in Fig.~\ref{fig:chanmaps32k} for the
32k data (taking into account the very different channel widths). Here.
we show only every second channel.

A comparison with the 4k channel maps in Fig.~\ref{fig:chanmaps4k}
  shows a marked increase in the amount of detail in the 32k data. A
  comparison is useful, as many wideband continuum studies with
  MeerKAT will  use the 4k correlator mode. The high sensitivity of
  MeerKAT will allow for the detection and characterisation of \HI in
  galaxies over a large fraction of the band, despite the wide
  channels in the 4k mode. Knowledge of how the observed
  characteristics of this \HI can differ between 4k and 32k data will
  thus contribute to a better understanding of the limitations of these 4k \HI data.

The channel maps in Fig.~\ref{fig:chanmaps32k} show details in the
tail (indicated in the 892.5 \kms channel map) and, together with
Figs.~\ref{fig:mk16} and \ref{fig:chanmaps4k}, establish that it winds
clockwise. There are also two features in the south. The one in the
south-east of the main body of the galaxy becomes visible starting
from the channel at 837.4 \kms (marked `1' in that channel) and
persists in the two subsequent channels. Its winding is
anti-clockwise, that is, opposite to that of the main feature in the
north. There is also a second feature in the south which becomes
clearly visible starting from the channel at 848.4 \kms (marked `2'
in that channel) and also persists in a couple of subsequent
channels. Its winding is clockwise, that is, the same as the main
northern feature. This feature can also be seen in
Figs.~\ref{fig:mk16} and \ref{fig:chanmaps4k} as well as in the
$r=1.5$ integrated \hi image given below.  We discuss the
possible nature of these features below. Finally, the faint clouds to
the south can also be seen in the 32k data, most clearly in the
channel map at 892.5 \kms.

\begin{figure}
  \includegraphics[width=0.90\columnwidth]{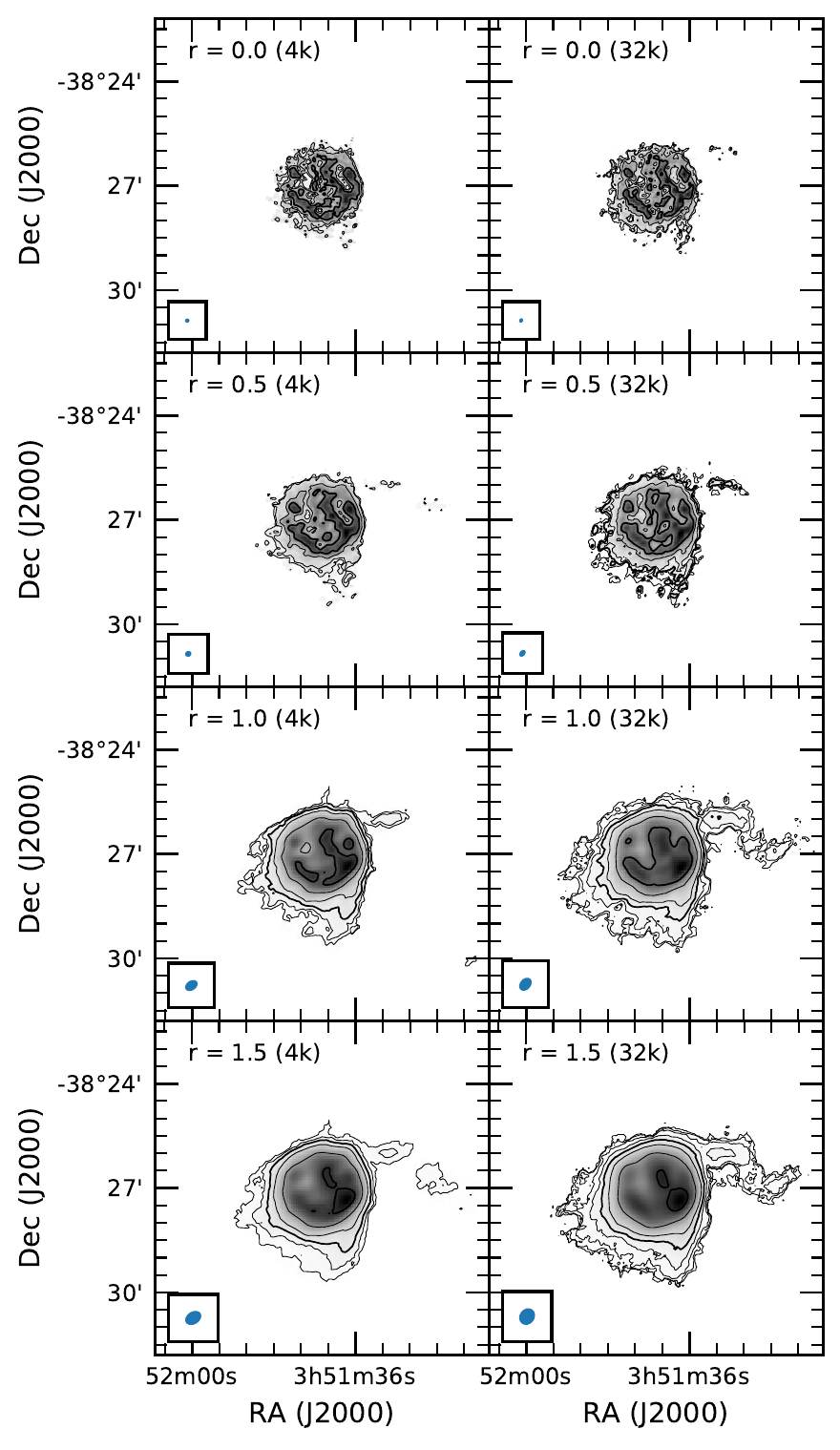}
  \caption{Integrated MeerKAT \HI (zeroth moment) maps of the galaxy ESO 302-G014 produced using
    different robust values. Rows show data imaged with different
    robust factors, given in the top-left of each panel, with values from top to bottom 
    $r= 0.0, 0.5, 1.0, 1.5$. Columns
    show the two data sets, with the 4k data on the left and the 32k
    data on the right.  The respective
    beams are indicated in the box in the lower-left corners of the
    panels, and are also listed in Table \ref{tab:noises}.
    The lowest contour levels shown are always the $3\sigma$ contour
    levels, with $\sigma$ given in Table \ref{tab:noises}, followed by
    contours at fixed approximately logarithmically-spaced column
    density levels.  This results in the following contour values (in
    units of $10^{20}$ cm$^{-2}$):  left
    column (4k data), from top to bottom:
    $r=0.0$: (2.9, 5.0, 10.0);
    $r=0.5$: (1.2, 2.0, 5.0, 10.0);
    $r=1.0$: (0.3, 0.5, 1.0, 2.0, 5.0, 10.0);
    $r=1.5$: (0.2, 0.5, 1.0, 2.0, 5.0, 10.0).
    Right column (32k data), from top to bottom:
    $r=0.0$: (1.2, 2.0, 5.0, 10.0);
    $r=0.5$: (0.4, 1.0, 2.0, 5.0, 10.0);
    $r=1.0$: (0.1, 0.2, 0.5, 1.0, 2.0, 5.0, 10.0);
    $r=1.5$: (0.06, 0.1, 0.2, 0.5, 1.0, 2.0, 5.0, 10.0).
    The $10^{20}$ cm$^{-2}$ and
    $10^{21}$ cm$^{-2}$ contours are drawn with slightly thicker lines.}
    \label{fig:mommaps}
\end{figure}

\subsection{Moment maps}

Figure \ref{fig:mommaps} shows the integrated \HI maps (zeroth-moment
maps) derived from the 4k and the 32k data for the four robust values
discussed here.  Small robust values emphasise 
high-resolution details in the disc, while the larger robust values (which are closer to natural weighting)
emphasise the more extended low-column density \HI distribution. 

The moment maps were created using the SoFiA source finder
\citep{serra_et_al_2015}. We again searched for emission brighter than
$3\sigma$ using various smoothing kernels. The 4k data cubes
were searched twice: once at their original resolution, and once with
a Gaussian spatial smoothing kernel with a full width at half maximum (FWHM) that is equal to the beam
size. Due to the limited number of channels no additional velocity smoothing
kernels were used here.

The 32k cubes were searched four times each using combinations of the
original spatial and velocity resolution, along with a Gaussian
spatial kernel equal to the beam size and a boxcar velocity smoothing
kernel of three channels. Larger kernels did not result in qualitatively
better maps.  The built-in comparison between positive and negative
detections with a reliability threshold of 0.90 was used for both data
sets to distinguish between noise peaks and real emission.
For all cases, we consider the column density limit of the
moment maps to be $3\sigma$ in one channel.

In Fig.\ \ref{fig:mommaps}, the 32k data reveal additional details not
seen in the 4k data set despite the somewhat shorter integration time
of the former (cf.\ Table \ref{tab:observations}).  This increased
column density sensitivity is due to the 32k channel width being
better matched to the width of the \hi profiles (typical FWHM values are  $\la 20$ \kms)
than the 4k channels.

The $r=0.0$ maps show a regular disc, with only slight indications of
the extension to the west. The disc becomes more and more asymmetric
and lopsided towards higher $r$ values. The $r=1.0$ and $r=1.5$ maps
clearly show the western filament, while towards the northern edge the column density contours are
much closer together than towards the southern edge. The change from a
regular disc to the more asymmetrical outer parts occurs around $\sim 1
\cdot 10^{20}$ cm$^{-2}$.

The western filament shows a number of kinks or breaks in its
morphology and extends further east than the moment maps suggest. The
channel maps between 870.5 and 903.6 \kms in
Fig.~\ref{fig:chanmaps32k} show that it connects with the
north-eastern part of the disc. In the moment maps this is, however,
hidden by superimposed bright \hi emission from lower
velocities. As the channel maps at 870.5 \kms and 881.5 \kms are also
the ones where the southern extension is most prominently visible, it
is possible that the western filament and the southern extension of
the disc are related.

\begin{figure}
  \includegraphics[width=0.95\columnwidth]{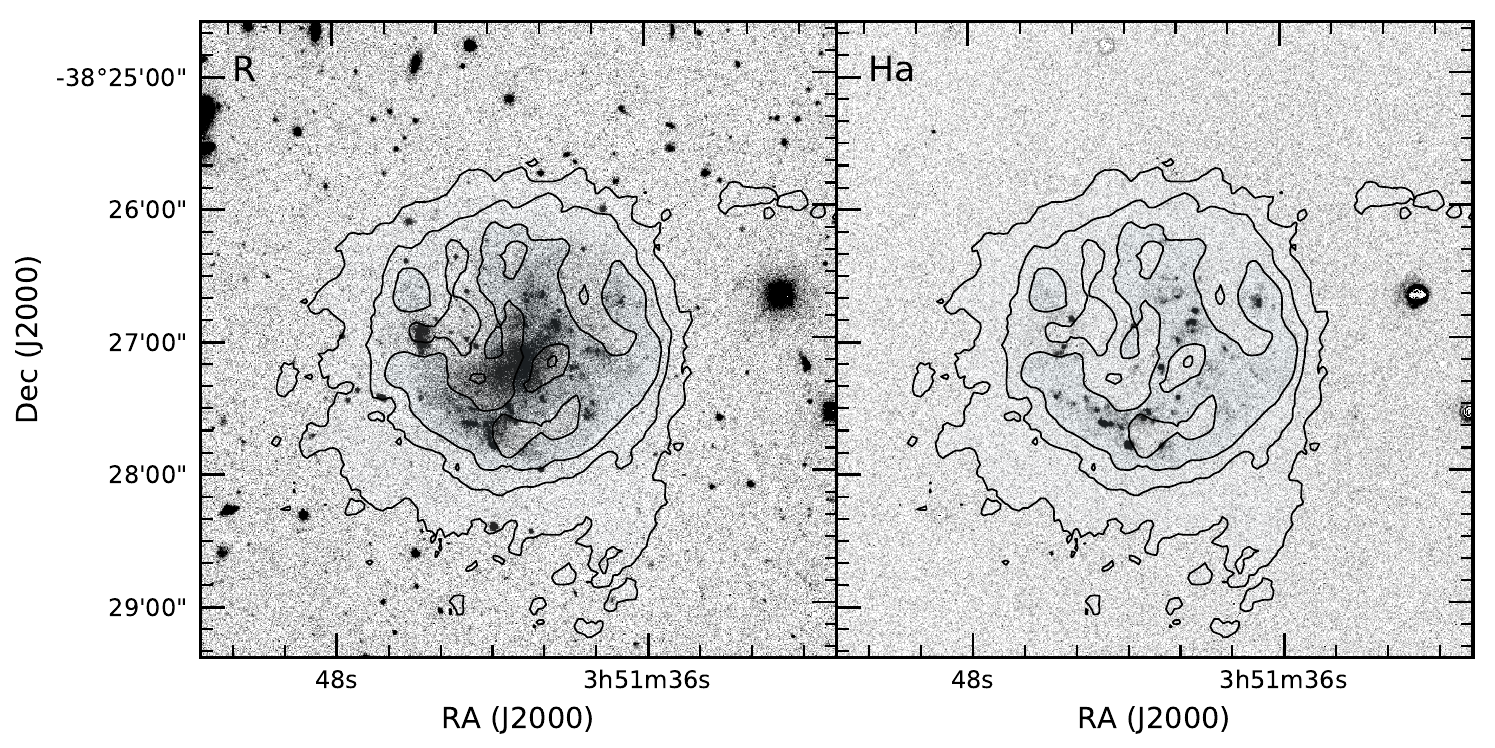}
  \caption{Overlays of the MeerKAT $r=0.5$ integrated \hi map on SINGG
    $R$-band (left panel) and H$\alpha$ (right panel) images. For
    clarity, only a limited number of column density contours are
    shown. These are $(1.0, 5.0, 10.0) \cdot 10^{20}$ cm$^{-2}$. Darker
    shading indicates the highest column densities.
    \label{fig:optical}}
\end{figure}

Figure~\ref{fig:optical} shows an overlay of the $r=0.5$ integrated
\hi map on the SINGG $R$-band and H$\alpha$ data. The filament and
southern extension extend well beyond the optical disc, as
expected. The $R$-band image shows no optical counterpart down to the
limiting surface brightness of 27.6 AB mag arcsec$^{-2}$. In the
H$\alpha$ image we also do not see evidence for recent star formation
at these locations. The two prominent young star clusters are located
at an under-density in the \hi. We will return to these clusters in
the next section. DECaLS images show a faint object close to the tip
of the filament at $(\alpha, \delta) = 03^h51^m20^s,\ -38^\circ
25'35''$. It appears in both $g$ and $r$ bands, so ensuring that the
object is real. We discuss this object further in
Sect.\ \ref{sec:interaction}.

\begin{figure}
  \includegraphics[width=0.95\columnwidth]{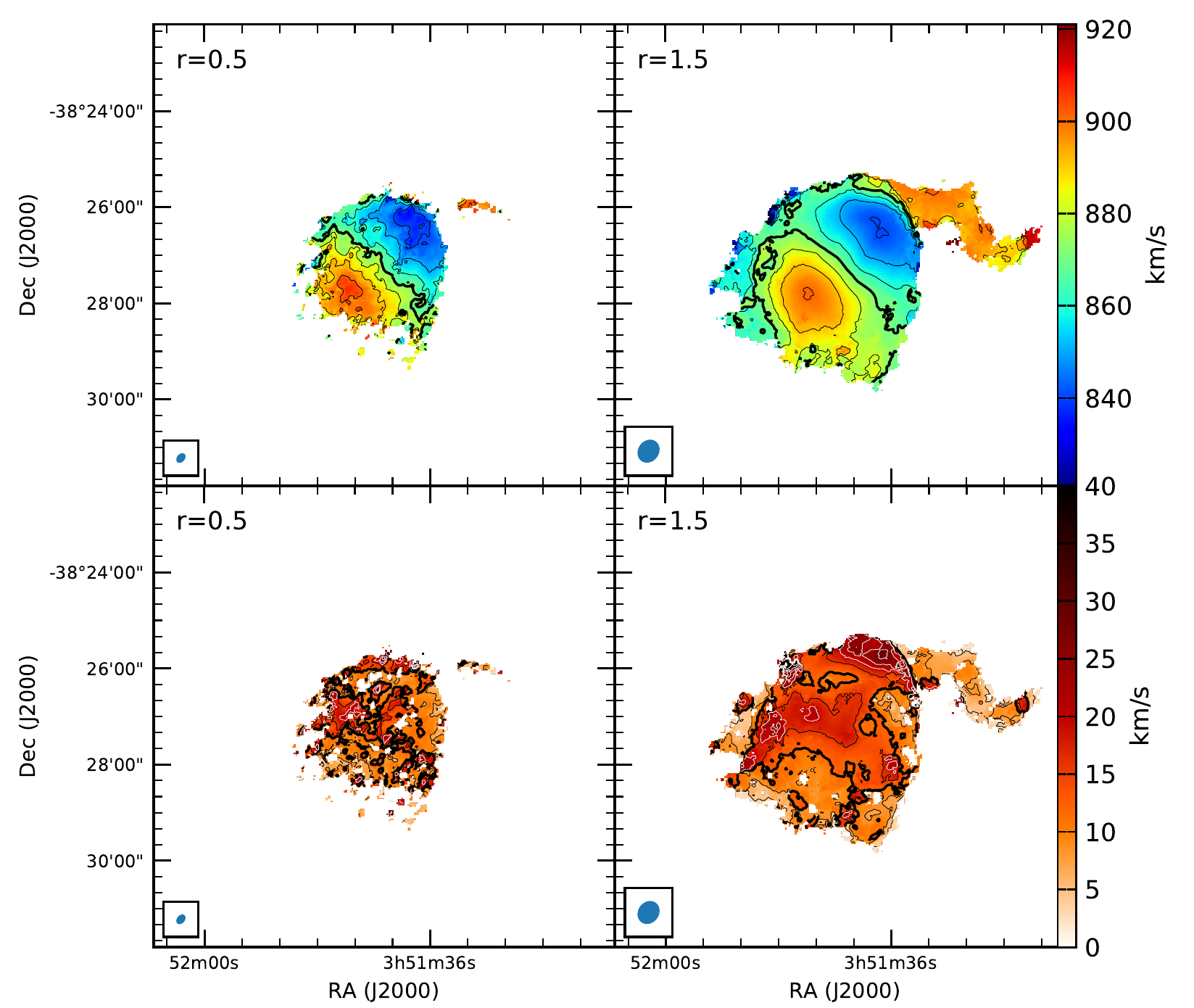}
  \caption{First- and second-moment maps of the galaxy ESO 302-G014 based on the MeerKAT 32k data. The
    top row shows the intensity-weighted mean velocity fields, the
    bottom row velocity dispersion maps.  The left column show the
    $r=0.5$ map, the right column the $r=1.5$ version.  An intensity
    cut has been applied to the maps. The left column shows only
    velocities where the column density is above $1\cdot 10^{20}$
    cm$^{-2}$. In the right column the limit is $1\cdot 10^{19}$
    cm$^{-2}$. In the top row the thick contour denotes a velocity of
    871 \kms (the systemic velocity). From there contours are shown in
    intervals of 10 \kms.  In the bottom row, the thick contour
    indicates 12 \kms. Contour values change in intervals of 4
    \kms. The highest values are indicated by white contours.
    \label{fig:velfi}}
\end{figure}

The intensity-weighted mean velocity fields (shown in
Fig.~\ref{fig:velfi}) paint a similar picture. The inner disc shows a
very regular overall velocity field, with only small ($\sim 5-10$ \kms)
perturbations (visible in the high-resolution data) that are
consistent with turbulence and star-formation related non-circular
motions.

The velocities of the extension differ significantly ($\sim$ 50-60
\kms) from those in the northern part of the disc and are more
consistent with those found in the southern part, potentially
hinting at a connection between the western filament and the southern
extension. 

The kinematic center seems to be offset from the center of the
bright \hi  emission. We note that the kinematic minor axis is not precisely perpendicular
to the kinematic major axis, suggesting the presence of a
non-axisymmetric structure in the main disc of the galaxy.
However, the photometric position angle of the disc is difficult to determine. There is a difference of $\sim$ 20$^\circ$ between the major axis position angles of the mid-infrared Spitzer data at 3.6 
and 4.5 $\mu$m (cf.\ archival data of \citealt{munoz_mateos_et_al_2015}), with the latter being close to the kinematic position angle, but these images
are rather faint. Deeper near infrared imaging is necessary to get a better understanding of the structure of the stellar disc of this galaxy.

The second-moment map shows a region of high velocity dispersion in
the central part of the disc, extending towards the east. This
corresponds with the star formation regions just to the north of the
optical center and also includes the two young star clusters in the
eastern part of the disc. The high values at the northern and extreme
eastern edge of the disc correspond to locations where the filament
crosses the main body of the \hi disc. These values are, therefore, due to the double velocity profiles along the line of sight, rather than a physical velocity dispersion.

\subsection{Fluxes and total \hi masses}

\begin{figure}
  \centering\includegraphics[width=0.8\columnwidth]{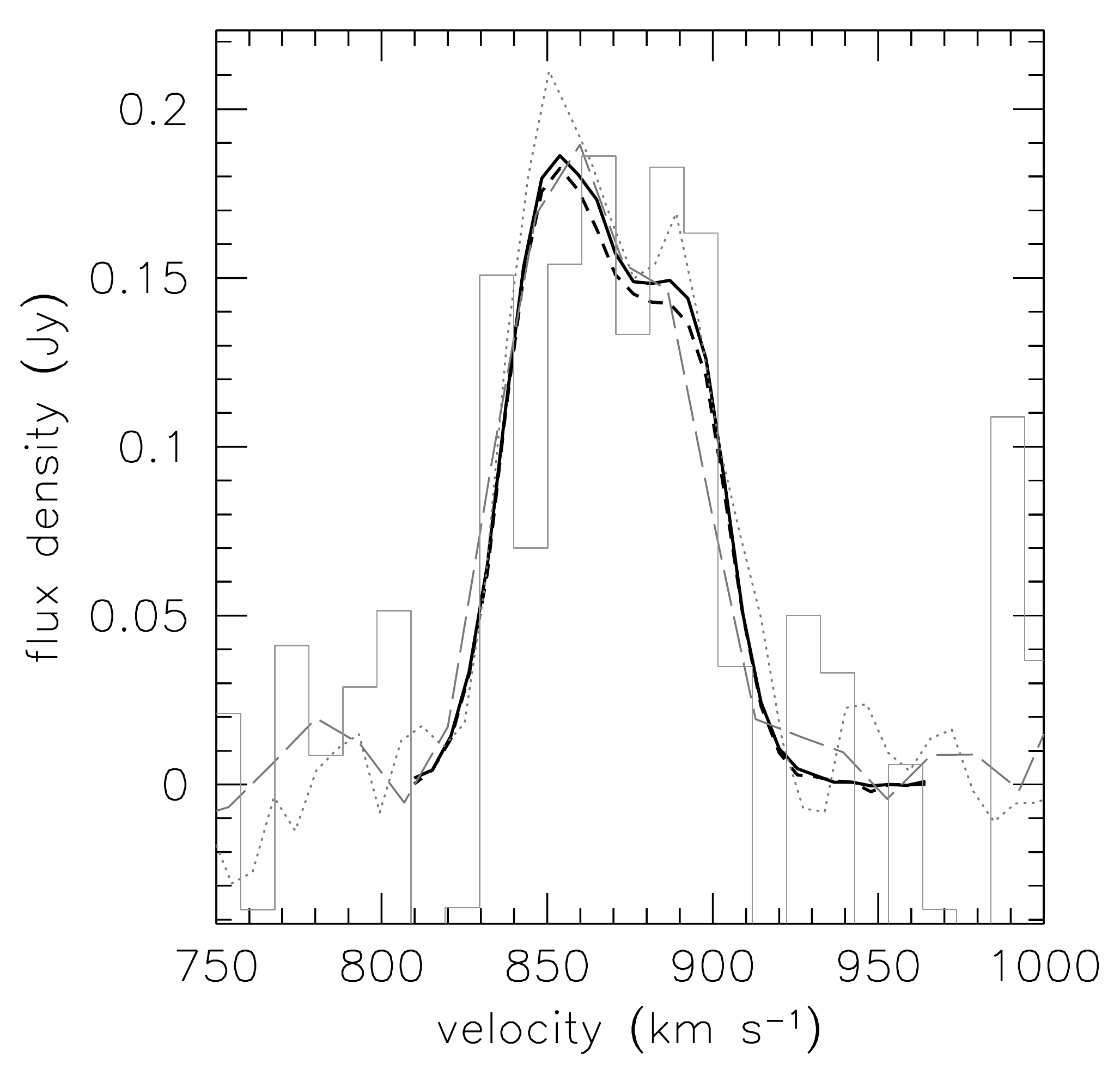}
  \caption{Integrated \hi and CO spectra (global profiles). The full curve
    shows the $r=1.5$ \hi profile, the dashed curve the $r=0.5$
    profile from MeerKAT. The dotted gray line is the \hi global profile derived from
    GBT data \citep{sardone_et_al_2020}. The long-dashed gray line is
    the HIPASS spectrum. The histogram indicates the global ALMA CO spectrum where the CO flux density values have been scaled down by a factor of three.\label{fig:globprof}}
\end{figure}

Table \ref{tab:noises} also lists the total fluxes derived from the
moment maps and the corresponding \HI masses.  The integrated flux
spectrum (global profile) is shown for the 32k data in
Fig.~\ref{fig:globprof} for two robust values. The highest robust
value recovers slightly more flux over the entire velocity range.

Uncertainties in the integrated fluxes and \hi masses were derived for the MeerKAT-16
and 32k data using the prescription given in
\citet{koribalski_et_al_2004}. We do not give
uncertainties for the 4k data due to the large channel width and
resulting small number of channels containing emission.

We can compare these values with the integrated fluxes (and \hi
mass) derived from single-dish observations. The flux value listed in the
HIPASS Bright Galaxy Catalogue is $11.2 \pm 2.0$ Jy \kms ($3.6 \cdot 10^8$
\msun) \citep{koribalski_et_al_2004}. This value is derived using the
HIPASS \hi position. For this work, we have gone back to the HIPASS
data and re-extracted the integrated spectrum using the optical
position.  We find a value of $11.4 \pm 0.2$ Jy \kms ($3.7 \cdot 10^8$
\msun). This updated spectrum is shown in Fig.~\ref{fig:globprof}.
We note the good agreement with the MeerKAT data.

ESO 302-G014 was also observed using the Green Bank Telescope (GBT). This observation 
was first presented in \citet{sorgho_et_al_2019}, where a flux of $10.4 \pm 0.4$ Jy \kms ($3.4 \cdot 10^8$ \msun) was derived. 
This was derived from masked data, so it should be considered a lower limit. 
These data were re-analysed in \citet{sardone_et_al_2020}, where a flux of 13.7 Jy \kms ($4.4 \cdot 10^8$ \msun) 
was found. This integrated spectrum is also shown in Fig.~\ref{fig:globprof}. In general, the agreement is good, 
except for the two peaks of the profile, where the GBT finds more flux. The angular size of the galaxy 
compared to the length and large number of short baselines of the MeerKAT data gives no reason for us to suspect 
that these are affected by  zero-spacing issues. Rather, as in terms of column-density sensitivity 
the GBT observations are close to an order of magnitude deeper, it is possible that these peaks indicate 
the presence of a few times $10^7$ \msun of additional \hi between $\sim 10^{17}$ and $\sim 10^{18}$ cm$^{-2}$. 
The full MHONGOOSE data set of this galaxy should be able to confirm the presence of this
low-column density \hi.

\subsection{Clouds}

The 4k observations showed the presence of faint \hi clouds $\sim 7'$
south of the main body of the galaxy in the central channel map. A
smoothed version of that channel map (left panel of
Fig.~\ref{fig:chanmapsmo}) shows this more clearly. This is based on
the $r=1.5$ 4k data smoothed to $45''$.

The complex is also visible in individual channel maps of the 32k
data; compare with the channel map at 892.5 \kms in
Fig.~\ref{fig:chanmaps32k}. Smoothing the 32k data also enhances the
visibility of this complex. The right-hand panel in
Fig.~\ref{fig:chanmapsmo} shows a zeroth-moment map based on $45''$,
$r=1.5$ 32k data, constructed using the same set of thresholds and kernels
as used for the moment maps shown in Fig.~\ref{fig:mommaps}.

The cloud is somewhat resolved ($\sim 2-3$ beams) but has a low column-density
overall. Its peak value reaches only $1.8 \cdot 10^{19}$ cm$^{-2}$ in
the $45''$ map. In the full-resolution, $r=1.5,$ moment map, the cloud
is not included due to the stringent criteria used. If
these are relaxed, the cloud is detected with a peak column density of
$4\cdot10^{19}$ cm$^{-2}$. Nevertheless, the reality of the cloud is
not in doubt due to its detection in the channel maps of two
independent data sets.

The mass of the cloud as detected in the $45''$ data is $9.3 \cdot
10^{5}$ \msun. This should be compared with the $45''$ resolution
total \hi mass of the galaxy of $3.8 \cdot 10^8$ \msun, so the cloud
contains only about 0.25\% of the total \hi  mass. In comparison, the \hi mass
of the western filament is $\sim 4.5 \cdot 10^6$ \msun, and given its
morphology it is possible this filament consists of a number of similar
low-mass clouds.

Figure~\ref{fig:chanmapsmo} also shows an overlay of the smoothed moment
map on a MeerLICHT $q$-band image. MeerLICHT is a wide-field 65-cm
optical telescope located at the South African Astronomical
Observatory in South Africa that observes any of MeerKAT's targets in
a number of optical bands simultaneously with the MeerKAT
observations. This overlay shows no optical counterparts for the
filament and cloud at least down to the surface brightness levels of
the E 302-G014 main disc.  The deeper DECaLS images show no evidence
for an optical counterpart of the cloud.

\section{Discussion}

The \HI moment maps in Fig.~\ref{fig:mommaps} show a changing
morphology for ESO 302-G014 when going from high to low column
densities. At high column densities the disc is regular and appears
almost unperturbed, with just a slight hint of the presence of the
western filament. This changes dramatically when column densities that are an
order of magnitude lower are probed. The western filament has become
more prominent and in comparing the northern and southern edges of the
disc, we can see  a northern disc edge that is more compressed than the
southern edge.

The central positions of the stellar and \HI components do not
coincide either. The centroid of the bright \HI emission (the axi-symmetric
inner part of the \HI disc) lies $\sim 12''$ or $\sim 680$ pc to the
north of the center of the stellar component (cf.\ left panel of
Fig.~\ref{fig:optical}).  

Lopsidedness or asymmetries are common in disc and dwarf
galaxies. They can manifest themselves as morphological lopsidedness,
but also as kinematical lopsidedness, or as asymmetries in the
integrated \HI spectra. For an extensive overview, see the review by
\citet{jog_et_al_2009} and references therein. \citet{jog_et_al_2009}
note that at least 30 percent of disc galaxies show a lopsided stellar
component, but that this frequency is higher in \HI. For example,
\citet{richter_et_al_1994} found that half of the 1700 galaxies they
studied showed a strong lopsidedness in their global \HI profiles. A
recent study by \citet{watts_et_al_2020} of 562 galaxies finds a
proportion of 37\%. \citet{espada_et_al_2011} studied
  lopsidedness in \hi and distinguished between isolated and field
  galaxies. They find that only $\sim 2\%$ of isolated galaxies show
  significant asymmetries, while this rate is much higher ($\sim$
  10-20\%) in regular field galaxies.

There have been many attempts to explain lopsidedness using a variety
of scenarios. Of the ones that are relevant here, the most promising
ones are minor tidal interactions, the accretion of gas from the
intergalactic medium, an off-centre disc within its halo, and ram
pressure stripping \citep{jog_et_al_2009}.
\citet{reynolds_et_al_2020} find a trend of increasing asymmetry
strength with density of the environment, as also indicated by
\citet{espada_et_al_2011}.  It is thus likely that ram pressure
stripping and tidal interactions are the most likely scenarios in the
higher density environments, with gas accretion explaining asymmetries
seen in more isolated galaxies.
We discuss a number of these scenarios and their relevance for ESO
302-G014 in more detail below.

\subsection{Ram pressure stripping scenario}

Looking at the large-scale environment of ESO 302-G014, we find that
(in projection) it is located in the vicinity of the Fornax cluster, at the northern
end of a large-scale structure filament that connects the Dorado group
with the Fornax cluster (see Fig.~1 in \citealt{venhola_et_al_2018}). 
The projected angular distance to the centre of the main component of
the Fornax cluster (NGC 1399) is $\sim 4.0^\circ$.

ESO 302-G014 is, however, significantly closer to us than Fornax and at 
the distance of ESO 302-G014 this projected distance corresponds to 0.8 Mpc.
Using the radial velocities of NGC 1399 (1425 \kms) and ESO302-G014 (871 \kms) as
distance indicators, we find a line-of-sight distance of 8.6
Mpc. For comparison, the virial radius of the Fornax cluster is 0.7 Mpc
\citep{drinkwater_et_al_2001}.  A recent catalogue of nearby groups
\citep{kourkchi_et_al_2017} confirms that ESO 302-G014 is not a member
of Fornax nor of any of the groups in the Fornax region.  It is,
therefore, unlikely that ram pressure stripping due to infall in Fornax
can explain the morphology of the galaxy.

It is, however, suggestive that the compression of the northern edge
of the disc points towards the Fornax cluster.  Is it therefore
possible that ram pressure due to the IGM between Fornax and the
Dorado group can cause the observed morphology?

Motivated by observations of NGC 300 in the Sculptor group,
\citet{westmeier_et_al_2011} show that for a galaxy moving at 200
\kms, the density of the IGM needs to be $\sim 10^{-5}$ cm$^{-3}$ for
ram stripping to occur, comparable with the density of the Local Group
medium.  This density is similar to the estimate by
\citet{bureau_et_al_2002}, who studied the effect of ram pressure on
the \hi distribution of Holmberg II, a dwarf galaxy falling into the
M81/M82/NGC3077 group. 

These two case studies show that the minimum
IGM densities needed for ram pressure stripping to occur are those
typically found in a group environment.
Simulations by \citet{pilkington_et_al_2011} of a Holmberg II-like
dwarf galaxy suggests that a lopsidedness with an apparent
`compression' on one side can also be caused by internal processes.
The presence of the western extension is also  not readily
explained by ram pressure stripping. In summary,  ram pressure stripping is, therefore, not expected  to play 
 a major role in defining the \hi morphology of
ESO 302-G014.

\begin{figure*}
  \centering
  \includegraphics[width=0.95\textwidth]{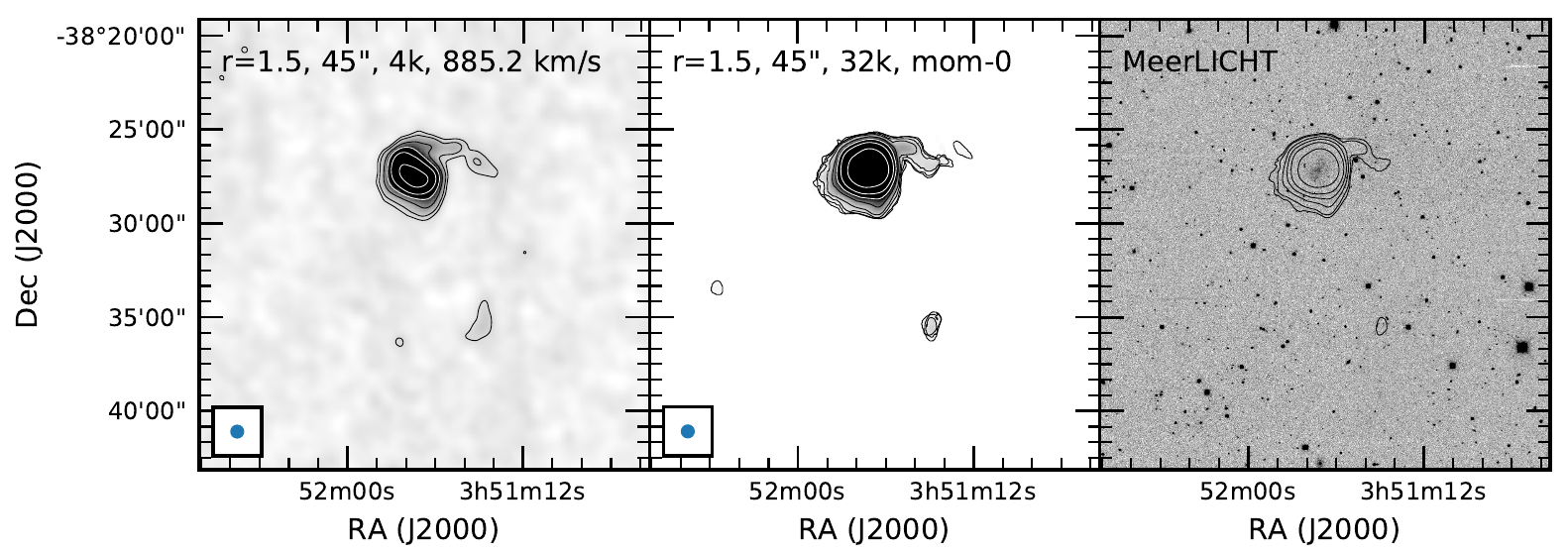}
  \caption{Lower resolution $45''$ maps based on the $r=1.5$
    data. Left panel: 4k data channel map at 885.2 \kms. Contour
    levels are (0.1, 0.2, 0.5, 1.0, 2.0, 5.0) $\cdot 10^{20}$
    cm$^{-2}$. Middle panel: 32k data integrated \hi (zeroth moment)
    map. Contour levels are (0.03 (3$\sigma$), 0.05, 0.1, 0.2, 0.5,
    1.0, 2.0, 5.0) $\cdot 10^{20}$ cm$^{-2}$. Right panel: MeerLICHT $q$-band optical image, with 32k integrated \hi map as overlay. Contour levels are (0.1, 0.2, 0.5,
    1.0, 2.0, 5.0) $\cdot 10^{20}$ cm$^{-2}$. }
    \label{fig:chanmapsmo}
\end{figure*}

\subsection{Interaction and accretion scenario\label{sec:interaction}}

Since ram pressure stripping is 
unlikely to explain the \hi morphology of the galaxy, we consider the possibility that its origin lies in the infall of a smaller galaxy.  The \hi mass
of the western filament as visible in the moment maps is $\sim 4.5
\cdot 10^6$ \msun. Taking into account that these integrated maps do
not show the eastern part of the extension, the total \hi mass of the
extension is about a factor of two larger. Assuming that all this \hi 
originated in the infalling galaxy, then this would have been a $\sim
10^7$ \msun dwarf galaxy.

As described in Section \ref{sec:description}, the features in the
northern part of the disc unwind in the same sense as the southwestern
one (feature 2 in Fig.\ref{fig:chanmaps32k}). These two together form
a two-armed spiral-like feature that would be driven by a small companion
or satellite. This could have either already merged, or be linked to
one or more of the gas clouds, or to some clump in the galaxy itself,
that is, on its way to a merging. In this case, the closeness of the
iso-density contours in the northern part of the galaxy would be
linked to the higher density of the northern arm, compared to the
southern one.

A second interesting feature in the data suggesting an interaction or
accretion scenario is the cloud to the south of ESO 302-G014.  It has
a central velocity of 892 \kms with a velocity width $W_{50} \simeq
25$ \kms. This recessional velocity agrees well with that of the
western filament (cf.\ Fig.~\ref{fig:chanmaps32k}).  The left panel in
Fig.\ \ref{fig:chanmapsmo} shows the filament curving south, and
extending this curvature, an arc can be drawn that connects the
filament with the cloud.  This suggests that all these features may be
related. A careful study of the MeerKAT data of this area did not show
any signs of the connecting \hi, but this may simply mean that it is
either of even lower column-density, or partly ionised. The full-depth
MHONGOOSE data should be able to constrain this much better.

The data are not deep enough or resolved enough to study the dynamics
of the cloud, but it is unlikely to be a dwarf galaxy.  Its \hi mass
is only slightly less than that of Leo T, the lowest mass gas-rich
galaxy currently known with an \hi mass of $4.1 \cdot 10^{5}$ \msun
\citep{adams_et_al_2018}. The peak column densities in the cloud are,
however, a factor of ten lower than those found in Leo T (even though
spatial resolutions are similar). The cloud measures $\sim 3.5$ kpc
along its minor axis, while Leo T has an \hi diameter of $\sim 0.8$
kpc measured at similar column density levels. The cloud is therefore
an altogether more diffuse entity than Leo T. Furthermore, neither the MeerLICHT
image shown in Fig.\ \ref{fig:chanmapsmo}, nor the DECaLS images show
evidence for an optical counterpart.

As noted above, the centers of the stellar distribution and that of
the inner (high column density) \HI disc do not coincide.  The
position of the center of the brighter parts of the \hi image is
offset with respect to the center of the stellar component (by $\sim
12'' = 680$ pc to the north).  Such a mismatch has been discussed in
(amongst others) a paper by \citet{pardy_et_al_2016}. They used
numerical simulations to study the interaction of an LMC-sized barred
irregular galaxy with a small companion with a mass ratio of 1:10, and
found that the dynamical center should coincide with the center of the
bar, rather than the center of the \hi distribution. While ESO
302-G014 shows no pronounced bar, the tidal effects we now witness may
have been due to a similar, earlier interaction with a much smaller,
dwarf companion, which by now has been captured, damping the amplitude
of the bar in the process (e.g. \citealt{athanassoula_1996}).

The deep DECaLS images mentioned in Sect.\ \ref{sec:auxdata} show
  a faint and diffuse object close to the tip of the filament with
  properties consistent with those of a dwarf galaxy
  (Fig.\ \ref{fig:hst}, left panel). A Sersic model fit gives an
  effective radius of 0.71 kpc, a Sersic index of 0.77, a $g-r$ color
  of 0.37 mag (no extinction correction applied), and an absolute
  magnitude of $M_g = -10.0$, all assuming that the object is at the same
  distance as ESO 302-G014. These properties are consistent with those
  of a diffuse dwarf galaxy. It is worth noting that the object is
  extremely faint, with a mean surface brightness within the effective
  radius in the $g$-band of $\langle \mu_g \rangle = 27.8$ mag
  arcsec$^{-2}$. Its slightly blue colours make it less likely that it
  is a background galaxy. While no \hi emission is seen at this
  position in the integrated \hi maps, inspection of the $r = 1.5$ 32k
  cube shows a $3\sigma$ peak close to the optical position but only
  in a single channel. The forthcoming deeper MHONGOOSE observations
  should show whether this \hi signal is real and whether we can
  associate this potential dwarf with ESO 302-G014. However, even if
  confirmed, it is not clear that this optical dwarf would be
  responsible for both the tail and the southern cloud. Thus, a more
  complex explanation would  still be needed.

Currently, the nearest major galaxy to ESO 302-G014 is ESO 302-G009 (HIPASS
J0347--38), at a projected distance of $\sim 180$ kpc. It is more
massive and slightly more distant (\citealt{meurer_et_al_2006} give a
distance of 13.4 Mpc for ESO 302-G009).  The deep GBT observations
presented in \citet{sorgho_et_al_2019} and \citet{sardone_et_al_2020}
show no evidence of \hi features between the two galaxies and it is
unlikely that tidal interactions between these two galaxies can cause
the more small-scale features that we observe around ESO 302-G014.

\citet{bournaud_et_al_2005} also studied lopsidedness in galaxies. They
found that (in the near-infrared) out of a sample of 149 disc galaxies,
two-thirds have significant lopsidedness. Late-type galaxies tend to
be more lopsided in their study. \citet{bournaud_et_al_2005} presented simulations that they
use to constrain the origins for lopsidedness. One of the conclusions
is that any tidal interaction or merger can only occur with a small
companion of a mass ratio of 10:1 or more. More massive companions
would transform the main galaxy towards an elliptical galaxy.

A second scenario, one explored by \citet{bournaud_et_al_2005}, is
asymmetrical accretion from a cosmic gas filament. They found that this
kind of accretion can result in strongly lopsided discs, even showing
arm-like features. A direct comparison with our data is not
straightforward, as the \citet{bournaud_et_al_2005} simulations model
the stellar population, not the \hi disc. However, as the \hi is generally a more sensitive tracer of interactions than the stellar
population (e.g. \citealt{chakrabarti_et_al_2011}), it is likely that the effects described in
\citet{bournaud_et_al_2005} should also be visible in the \hi,
and possibly even more pronounced.

\subsection{A potential link with star formation\label{sec:SF}}

\begin{figure*}
  \includegraphics[width=0.98\textwidth]{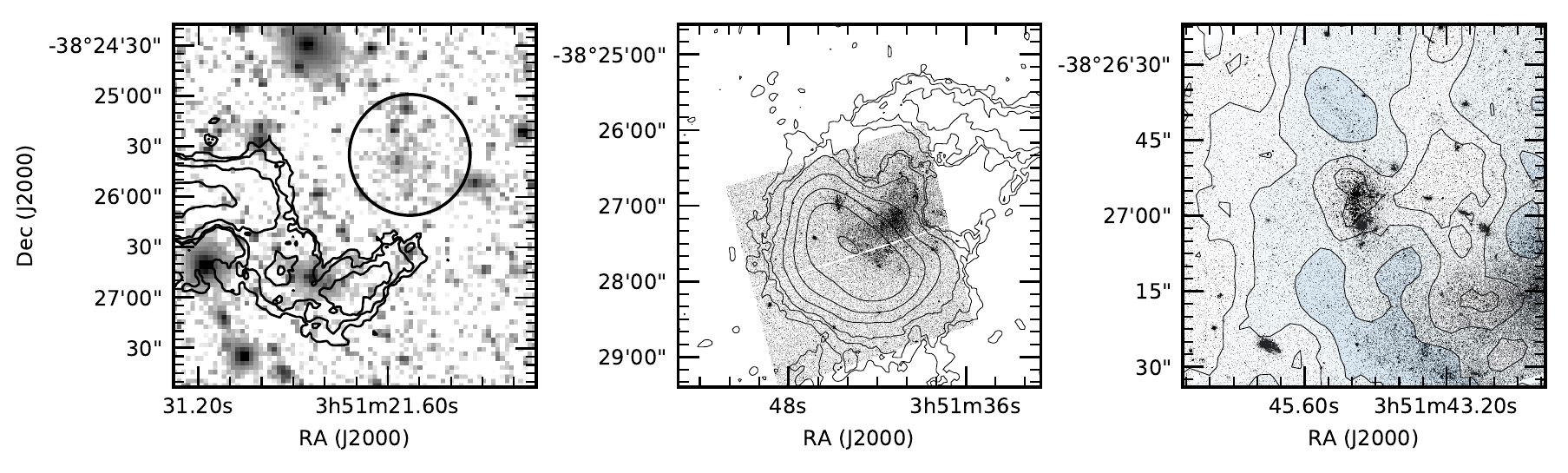}
  \caption{Left panel: Overlay of the $r=1.5$, 32k integrated \hi map on a combined $g+r$ image from DECaLS, binned to a pixel size of 2.7$''$ to increase signal-to-noise.
    Contour levels are as in Fig.\ \ref{fig:mommaps}. The circle indicates the dwarf galaxy candidate.
    Middle panel: Contours show the zeroth-moment map of
    emission $>2\sigma$ in the three channels between 892.5 and 903.6
    \kms in the $r=0.5$ 32k cube. Levels are $(2.5, 5, 10, 25, 50,
    100, 250)\cdot\sigma_{\rm mom}$, where $\sigma_{\rm mom} = 1.3
    \cdot 10^{19}$ cm$^{-2}$.  Right panel: Zoom-in of the HST image
    centered on the stellar clusters. Contours represent the $r=0.0$
    zeroth-moment map. Contour levels are $(2.0,5.0,10.0)\cdot
    10^{20}$ cm$^{-2}$.
    \label{fig:hst}}
\end{figure*}

As noted above, the integrated \hi maps do not show the full extent of
the western filament, as its eastern part is hidden by the bright
emission of main-disc \hi at different velocities. We created an
integrated \hi map of the emission between 892.5 and 903.6 \kms,
including all emission $>2 \sigma$ in each of the three channel maps.
This map is shown in Fig.\ \ref{fig:hst}, where it is overlaid on one
of the HST images discussed in Sect.\ \ref{sec:auxdata}.

\begin{figure*}
  \centering
  \includegraphics[width=0.95\textwidth]{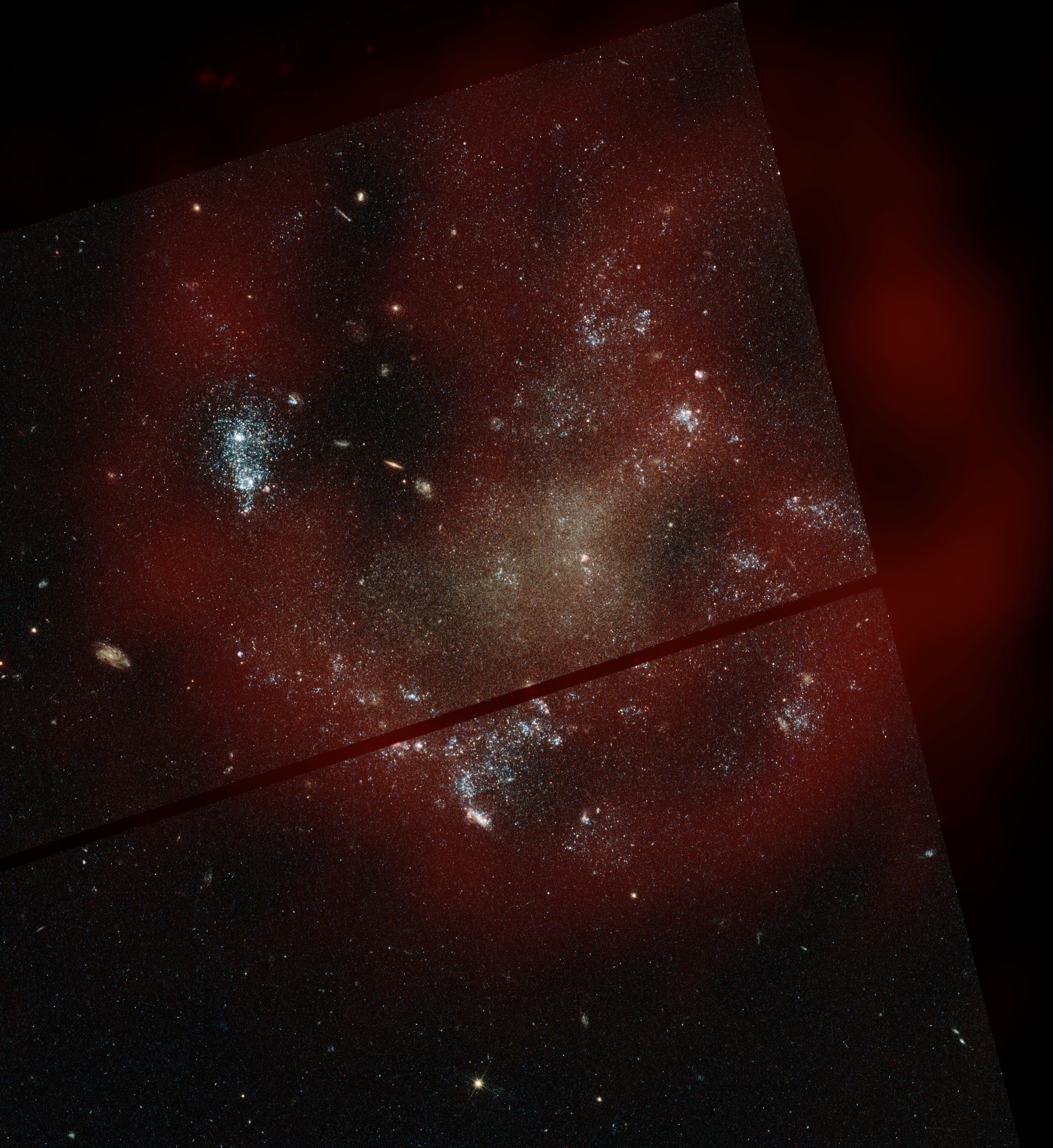}
  \caption{Combination of the MeerKAT $r=0.5$ integrated \hi map (red) overlaid on a colour combination of HST WFC3 IR F160W (transparent red), HST ACS F658N H$\alpha$ (red), HST ACS F606W (green-yellow), and HST ACS F336W (green-blue).
    \label{fig:hstcolor}}
\end{figure*}

The moment map shows the filament connecting with the main disc \hi
just to the north of two stellar clusters, which are the two brightest
stellar condensations in the galaxy. These clusters are shown in more
detail in the right panel of Fig.\ \ref{fig:hst}, where we have
overlaid the highest resolution $r=0.0$ integrated \hi map. 
An alternative presentation is shown in Fig.~\ref{fig:hstcolor}, which combines the MeerKAT image with the available HST imaging.

Both figures show 
a pronounced under-density in \hi at the position of the
clusters. This is the only place in the galaxy where we can find such
prominent stellar clusters. One possible explanation is that that this
star formation is triggered by the filament connecting it to the main
disc. \citet{prieto_et_al_2012} studied these clusters in detail and
derived an age for these clusters of $\sim 9$ Myr, with the
corresponding ultra-violet-based SFR accounting for $\sim 30$ \%
of the total SFR of the galaxy.

The possibility of enhanced star formation is also suggested by the
detection of molecular gas. Using the ALMA data described in
Sect.\ \ref{sec:auxdata}, we find an integrated CO$(J=1-0)$ flux
$S({\rm CO})\,dV$ of $32 \pm 5$ Jy km/s.
The integrated CO profile is shown in Fig.\ \ref{fig:globprof}. Using Eq.\ 3 from
\citet{solomon_et_al_2005} and a standard CO-to-H$_2$ conversion
factor, we find the molecular gas mass to be $4.6 \cdot 10^7\,M_\odot$,
which is significant for a low-mass late-type IB(s)m galaxy with a
luminosity of  $\sim 10^8\, L_\odot$. The molecular gas mass
almost equals the stellar mass, which is not commonly found in dwarf
galaxies. However, the metallicity is low in these kind of low-mass
galaxies, and the conversion factor is expected to be larger
\citep[e.g.][]{bolatto_et_al_2013}, resulting in a larger molecular mass.
The molecular to atomic mass ratio is $\sim 0.12$ and is
small compared to that found in spiral galaxies, but it is on the high
end of the range found in dwarf irregular galaxies
\citep{leroy_et_al_2009, young_et_al_1991}. The similarity of the CO
and \hi global profiles suggests the presence of molecular gas
throughout most of the disc.

It is therefore conceivable that the western extension is related to a
minor interaction or merger, leading to the currently observed star
formation and the significant molecular component along with contributing to the
lopsidedness or asymmetry of the galaxy. More examples of the
triggering of star formation in dwarf galaxies by minor interactions
are discussed in \citet{lelli_et_al_2014}.  \citet{segupta_et_al_2012}
also discuss a case similar to that of ESO 302-G014.  Their target
galaxy, UGC 1547 (CIG 85), is an irregular galaxy, with one edge more
compressed than the other, both in the \hi and optical. It has a regular
velocity field in the inner parts, is gas-rich and shows a recent
increase in star formation. In this case, \citet{segupta_et_al_2012}
identify the potential remnants of the two dwarf galaxies responsible,
showing that accretion of one or multiple satellites can indeed
explain the enhanced star formation and asymmetries.

\begin{figure}
    \includegraphics[width=\columnwidth]{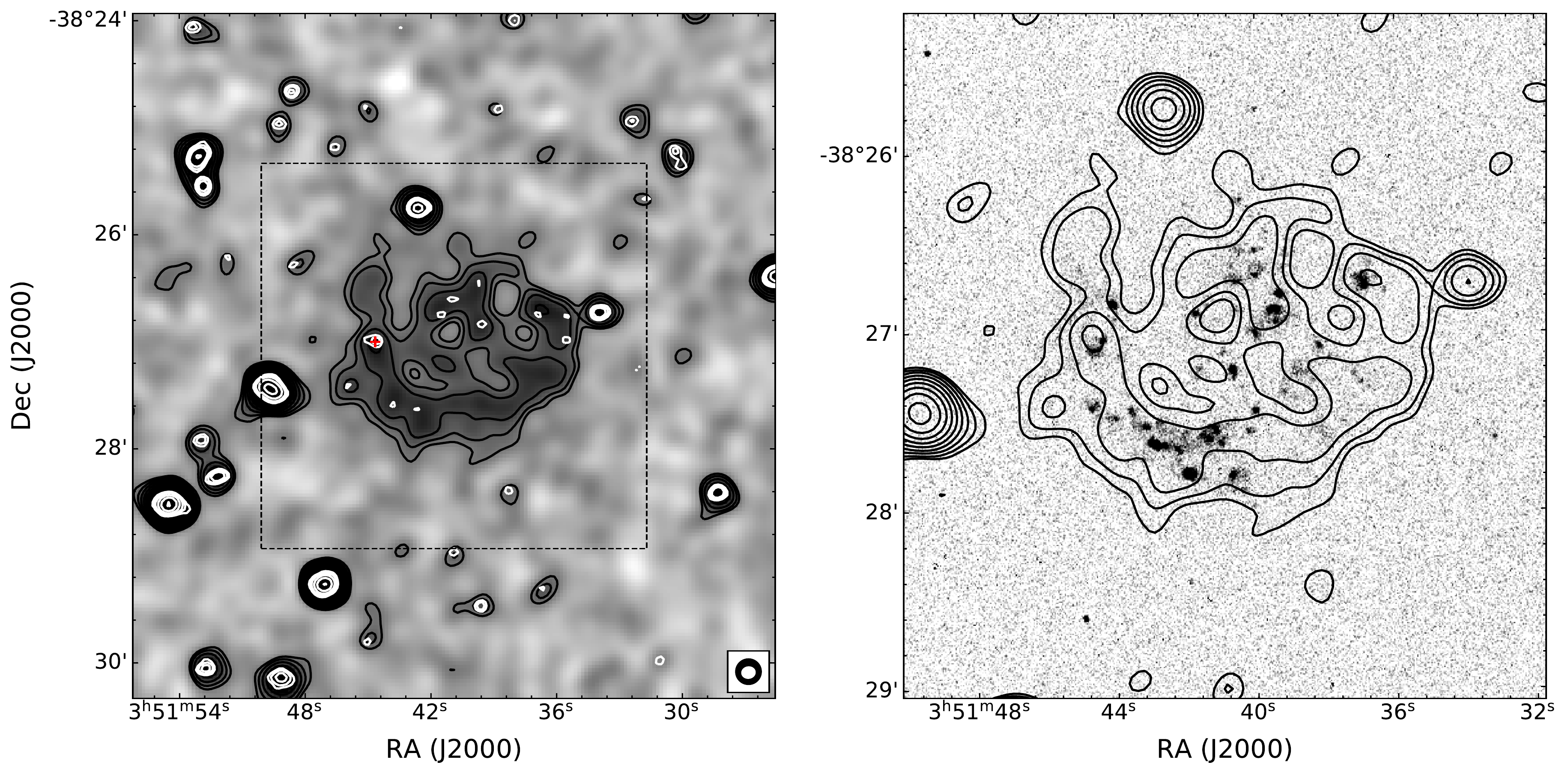}
    \caption{Comparison between radio continuum and H$\alpha$ emission. Left panel: Grayscale shows the radio continuum image smoothed to $15''$ resolution, and the black contours are from the same map starting at 3$\sigma$ and increasing by powers of $\sqrt{2}$. White contours are for the full-resolution continuum image starting at 4$\sigma$ and increasing by powers of $\sqrt{2}$ up to 8$\sigma$ and then by powers of two thereafter. The beam sizes of the radio continuum images are shown in the bottom right corner, with colors corresponding to their contours. The position of SN2008jb is shown with a red cross. Right panel: H$\alpha$ image from \citet{meurer_et_al_2006} in grayscale, within the region indicated by the dashed square in the left panel, and overlaid with the same $15''$-resolution radio continuum contours as in the left panel.}
    \label{fig:continuum}
\end{figure}

Further insights into the star formation history can be gleaned from the
radio continuum image shown in Figure~\ref{fig:continuum}. The
morphology is most clearly traced at $15''$ ($\sim$1~kpc)
resolution. Here, we can see a ridge of prominent radio continuum
around the periphery of the disc. The radio continuum emission is a
variable combination of thermal and non-thermal (synchrotron)
radiation, useful together as an indirect tracer of star formation;
using the SFR calibration from \citet{murphy_etal_2011}, we estimate
$\mathrm{SFR_{1.4GHz}=0.09\,M_\odot\,yr^{-1}}$ (corrected to a
Salpeter IMF). This is higher than the H$\alpha$ and UV SFR discussed
earlier, but this difference is most likely explained by the different
star-formation time scales involved and by ignoring of dust-obscured
star formation. An in-depth discussion of this SFR value is beyond the
scope of this paper.

In most locations, we see an excellent correspondence between the bright
continuum emission and H$\alpha$ features, suggesting that the thermal
fraction in those regions is relatively high. However, in the
southwestern region, the radio continuum ridge diverges from the arc
of H$\alpha$-emitting regions.  In that area, the radio continuum is
more synchrotron-dominated. The characteristic timescale for
synchrotron radiation at this frequency, for a typical magnetic field
strength of $5~\mu\mathrm{G}$, is $10^8$~yr \citep{condon_1992}. These
morphological features may indicate that star formation has progressed
around the periphery of the disc, clockwise from the southwest region
and proceeding most recently to the star-forming regions near
SN2008jb. We consider whether such a sequential progression of star formation around
the edge of the disc could have been spurred by the infall of an \hi\ filament
like the one currently seen to the west of ESO~302-G014. We anticipate
that this picture can be tested further by mapping the continuum
emission at another frequency and examining the variation in
synchrotron spectral index, which is indicative of the characteristic
age of the cosmic rays, in addition to carefully modelling the
\hi\ kinematics.

Finally, it is possible that some of the low-column density gas that
we find is due to direct cold accretion from the
IGM. \citet{putman_et_al_2012} suggest that the accreting gas is
mostly ionised and only condenses into neutral gas close to the
disc. At first glance, this seems consistent with the western
filament. However, the densest parts of the filament have column
densities of $\sim 5 \cdot 10^{19}$ cm$^{-2}$, meaning these kind of
clouds would have been detected at the sensitivity of HALOGAS. If
accretion from the IGM into such high column density features is
common, HALOGAS would have detected many more clouds.  The fact that
it has not (Kamphuis et al.\ in prep.)  suggests that any pure
accretion must take place at lower column densities and that the
western filament is indeed better explained by a minor
interaction or merger scenario.

\section{Summary}

In this paper, we present three MeerKAT \hi commissioning observations of
MHONGOOSE galaxy ESO 302-G014. At the highest column densities, the disc is regular and
symmetric. Going to lower column densities, the disc becomes more
lopsided and asymmetric.  The northern edge of the disc looks
compressed, while the southern edge is more diffuse.

The galaxy also exhibits an extension or tail of low-mass \hi clouds,
which could be connected with an isolated \hi cloud $\sim 7'$ ($\sim
23$ kpc) to the south of the main galaxy.
The tail and clouds could be the result of a minor interaction with a
$\sim 10^7$ \msun dwarf galaxy, which could also explain the
lopsidedness of the disc. Deep optical images show a faint potential
dwarf galaxy near the tip of the tail, but a confirmation of its
association with ESO 302-G014 will require deeper \hi observations.

The galaxy has a significant molecular component.  It is possible that
this, and the presence of prominent stellar clusters, is related to a
star formation event triggered by the interaction.  The final, deeper
MHONGOOSE observations should help explain some of these intriguing
features.

These MHONGOOSE commissioning observations have shown that MeerKAT
produces \hi data of exquisite quality. Here, we highlight the
capability to produce high-quality imaging over a large range in
angular resolution. This is a great advantage that will help provide a more complete observational picture of the fate of neutral
gas in galaxies as it moves from the IGM into galaxies.

%

%

\begin{acknowledgements}
It is a pleasure to thank the MeerKAT commissioning team without whom this paper would not have been  possible. 

The MeerKAT telescope is operated by the South African Radio Astronomy Observatory, which is a facility
of the National Research Foundation, an agency of the Department of Science and Innovation.

(Part of) the data published here have been reduced using the CARACal
pipeline, partially supported by ERC Starting grant number 679627
``FORNAX'', MAECI Grant Number ZA18GR02, DST-NRF Grant Number 113121
as part of the ISARP Joint Research Scheme, and BMBF project 05A17PC2
for D-MeerKAT. Information about CARACal can be obtained online under
the URL: {\tt https://caracal.readthedocs.io}.  This work is partly
based on data obtained with the MeerLICHT telescope, located at the
SAAO Sutherland station, South Africa. The MeerLICHT telescope is run
by the MeerLICHT consortium, on behalf of Radboud University, the
University of Cape Town, the Netherlands Foundation for Scientific
Research (NWO), the National Research Facility of South Africa through
the South African Astronomical Observatory, the University of Oxford,
the University of Manchester and the University of Amsterdam.  The
work of PK is partially supported by the BMBF project 05A17PC2 for
D-MeerKAT.  This project has received funding from the European
Research Council (ERC) under the European Union’s Horizon 2020
research and innovation programme (grant agreement no.\ 679627;
project name FORNAX). BKG acknowledges the UK’s Science \& Technology
Facilities Council (STFC), through the University of Hull Consolidated
Grant (ST/R000840/1).  KS acknowledges funding from the Natural
Sciences and Engineering Council of Canada (NSERC). AS is supported by
an NSF Astronomy and Astrophysics Postdoctoral Fellowship under award
AST-1903834.  EA and AB thank the CNES for financial support.  The
work of DJP was partially supported by NSF CAREER grant
AST-1149491. FB acknowledges funding from the European Union’s Horizon
2020 research and innovation programme (grant agreement
no.\ 726384/EMPIRE). LVM and JR acknowledge financial support from the
grants AYA2015-65973-C3-1-R and RTI2018-096228- B-C31 (MINECO/FEDER,
UE), as well as from the State Agency for Research of the Spanish MCIU
through the ``Center of Excellence Severo Ochoa'' award to the
Instituto de Astrofísica de Andalucía (SEV-2017-0709).

\end{acknowledgements}

\bibliographystyle{aa} 
\bibliography{mhon_e302} 

\begin{appendix}

\section{MHONGOOSE sample}

In Table \ref{tab:sample} we list some basic properties of the MHONGOOSE sample galaxies. A full description of the sample and the selection procedure will be the topic of a future paper.
\begin{table*}
    \caption[]{MHONGOOSE sample}
    \label{tab:sample}
    \centering
    \begin{tabular}{l l l l r r  r r}
      \hline
      \hline
      HIPASS &  Name  &      $\alpha$ (J2000.0)& $\delta$ (J2000.0)   &     $D$   &  $V_{\rm hel}$ & $\log(\frac{M_{\rm HI}}{M_{\odot}})$ & $W_{50}$\\
             &        &      $(^h\ ^m\ ^s)$   &   $(^\circ\ '\ '')$    &     (Mpc)   &  (\kms)  &  & (\kms)\\
      (1)& (2)& (3)& (4)& (5)& (6) & (7)& (8)\\
      \hline
      \multicolumn{8}{l}{Bin: $6.0<\log(M_{\rm HI}/M_{\odot})< 8.0$}\\
      \hline
      J0008--34   & ESO349-G031& 00 08 13.36  &--34 34 42.0  &3.3  &221.0  &               7.17  & 30  \\
      J0049--20   & UGCA015    & 00 49 49.20  &--21 00 54.0  &3.3  &294.6  &               6.99  & 36  \\
      J0310--39   & ESO300-G016& 03 10 10.48  &--40 00 10.5  &9.3  &709.7  &               7.95  & 36  \\
      J0454--53   & NGC1705    & 04 54 13.50  &--53 21 39.8  &5.1   &633.3 &               7.96  & 128 \\
      J1321--31   & KK98-195   & 13 21 08.20  &--31 31 45.0  &5.2  &571.8  &               7.56  & 38  \\
      \hline
      \multicolumn{8}{l}{Bin: $8.0<\log(M_{\rm HI}/M_{\odot})< 8.5$}\\
      \hline
      J0031--22   & ESO473-G024& 00 31 22.51  &--22 45 57.5  &7.9  &539.2  &               8.01  & 47  \\
      J0135--41   & NGC0625    & 01 35 04.63  &--41 26 10.3  &4.1  &395.7  &               8.09  & 75  \\
      J0320--52   & NGC1311    & 03 20 06.96  &--52 11 07.9  &7.0  &569.1  &               8.25  & 81  \\
      J0429--27   & NGC1592    & 04 29 40.13  &--27 24 30.7  &13.0  &946.0  &              8.37  & 73  \\
      J1337--28   & ESO444-G084& 13 37 19.99  &--28 02 42.0  &4.6  &587.0  &               8.03  & 58  \\
      \hline
      \multicolumn{8}{l}{Bin: $8.5<\log(M_{\rm HI}/M_{\odot})< 9.0$}\\
      \hline
      J0331--51   & IC1954     & 03 31 31.39  &--51 54 17.4  &14.3 &1063.4 &               8.96  & 224 \\
      J0351--38   & ESO302-G014& 03 51 40.90  &--38 27 08.0  &11.7 &871.7  &               8.55  & 67  \\
      J1106--14   & KKS2000-23 & 11 06 12.00  &--14 24 25.7  &12.7 &1039.3 &               8.62  & 77  \\
      J1253--12   & UGCA307    & 12 53 57.29  &--12 06 21.0  &8.6  &824.2  &               8.67  & 72  \\
      J2009--61   & IC4951     & 20 09 31.77  &--61 51 01.7  &11.3 &814.4  &               8.87  & 122 \\
      \hline
      \multicolumn{8}{l}{Bin: $9.0<\log(M_{\rm HI}/M_{\odot})< 9.5$}\\
      \hline
      J0309--41   & ESO300-G014& 03 09 37.87  &--41 01 49.7  &12.9 &955.0  &               9.00   & 130 \\
      J0546--52   & NGC2101    & 05 46 24.17  &--52 05 18.7  &16.1 &1192.3 &               9.18  & 93  \\
      J1303--17b  & UGCA320    & 13 03 16.74  &--17 25 22.9  &7.7  &742.9  &               9.12  & 113 \\
      J1318--21   & NGC5068    & 13 18 54.81  &--21 02 20.8  &6.9  &668.1  &               9.16  & 73  \\
      J2357--32   & NGC7793    & 23 57 49.83  &--32 35 27.7  &3.9  &227.3  &               9.01  & 173 \\
      \hline
      \multicolumn{8}{l}{Bin: $9.5<\log(M_{\rm HI}/M_{\odot})< 10.0$}\\
      \hline
      J0335--24   & NGC1371    & 03 35 01.34  &--24 55 59.6  &20.4 &1462.9 &               9.84  & 391 \\
      J0459--26   & NGC1744    & 04 59 57.80  &--26 01 20.0  &10.0 &740.8  &               9.56  & 200 \\
      J0516--37   & ESO362-G011& 05 16 38.80  &--37 06 09.1  &18.7  &1344.0 &              9.68  & 280 \\
      J1103--23\tablefootmark{a}& NGC3511    & 11 03 23.77  &--23 05 12.4  &14.2 &1113.9 & 9.62  & 265 \\
      J1254--10a  & NGC4781    & 12 54 27.00  &--10 30 30.0  &16.1 &1260.1 &               9.55  & 233 \\
      \hline
      \multicolumn{8}{l}{Bin: $10.0<\log(M_{\rm HI}/M_{\odot})< 10.5$}\\
      \hline
      J0052--31   & NGC0289    & 00 52 42.36  &--31 12 21.0  &22.9  &1628.9 &              10.34 & 277 \\
      J0419--54   & NGC1566    & 04 20 00.42  &--54 56 16.1  &20.7 &1503.0 &               10.19 & 205 \\
      J0445--59   & NGC1672    & 04 45 42.50  &--59 14 49.9  &18.1 &1334.0 &               10.19 & 250 \\
      J1153--28   & UGCA250    & 11 53 24.06  &--28 33 11.4  &24.4 &1706.5 &               10.00  & 274 \\
      J2257--41   & NGC7424    & 22 57 18.37  &--41 04 14.1  &13.6 &938.5  &               10.04 & 159 \\
        \hline
    \end{tabular}
    \tablefoot{ \tablefoottext{a}{pair with NGC 3513.} Columns: (1)
      HIPASS identification. (2) Other name. (3) Right ascension
      (J2000.0). (4) Declination (J2000.0). (5) Distance (Mpc). (6)
      Heliocentric velocity (\kms). (7) Logarithmic \HI mass. (8)
      $W_{50}$ \hi velocity width. Galaxy properties from
      \citet{meurer_et_al_2006}.  }
\end{table*}

\end{appendix}

\end{document}